\PassOptionsToPackage{pdfpagelabels=false}{hyperref}
\documentclass[fleqn,usenatbib]{mnras}

\usepackage[T1]{fontenc}

\usepackage{graphicx}	% Including figure files
\usepackage{amsmath}	% Advanced maths commands
\usepackage{amssymb}	% Extra maths symbols
\usepackage{newtxtext,newtxmath}

\def \kmsmpc      {km\,s$^{-1}$\,Mpc$^{-1}$}

\def \deg         {\text{$^{\circ}$}}

\def \arcsec      {\text{$^{\prime\prime}$}}

\def \mjybeam     {mJy\,beam$^{-1}$}
\def \mujybeam    {$\mathrm{\mu}$Jy\,beam$^{-1}$}
\def \klambda     {k$\lambda$}
\def \mach	  {\mathcal{M}}

\newcommand{\beam}[2]{{#1}\arcsec$\times${#2}\arcsec}
\newcommand{\beamtable}[2]{{#1}$\times${#2}}

\newcommand{\Msun}{\text{$\rm M_\odot$}}

\def \target {PSZ2\,G096.88+24.18}
\title[Radio relics in PSZ2\,G096.88+24.18]{Radio relics in PSZ2\,G096.88+24.18: A connection with pre-existing plasma}

\author[A. Jones et al.]{
A. Jones,$^{1}$\thanks{E-mail: ajones@hs.uni-hamburg.de}
F. de Gasperin,$^{1,2}$
V. Cuciti,$^{1}$
D. N. Hoang,$^{1}$
A. Botteon,$^{3}$
M. Br\"uggen,$^{1}$
\newauthor
G. Brunetti,$^{2}$
K. Finner,$^{4}$
W. R. Forman,$^{5}$
C. Jones,$^{5}$
R. P. Kraft,$^{5}$
T. Shimwell$^{3,6}$
\newauthor
and R. J. van Weeren$^{3}$
\\
$^{1}$Hamburger Sternwarte, Universit\"at Hamburg, Gojenbergsweg 112, 21029, Hamburg, Germany\\
$^{2}$INAF - Istituto di Radioastronomia, via P. Gobetti 101, Bologna, Italy\\
$^{3}$Leiden Observatory, Leiden University, PO Box 9513, 2300 RA Leiden, The Netherlands\\
$^{4}$Department of Astronomy, Yonsei University, 50 Yonsei-ro, Seoul 03722, Korea\\
$^{5}$Harvard-Smithsonian Center for Astrophysics, 60 Garden Street, Cambridge, MA 02138, USA\\
$^{6}$Netherlands Institute for Radio Astronomy (ASTRON), P.O. Box 2, 7990 AA Dwingeloo, The Netherlands
}

\date{Accepted XXX. Received YYY; in original form ZZZ}

\pubyear{2020}

\begin{document}
\label{firstpage}
\pagerange{\pageref{firstpage}--\pageref{lastpage}}
\maketitle

% Abstract of the paper
\begin{abstract}
Giant radio relics are arc-like structures of diffuse, non-thermal synchrotron radiation that trace shock waves induced by galaxy cluster mergers. The particle (re-)acceleration mechanism producing such radio relics is unclear. One major open question is whether relics can be formed directly from a population of thermal seed electrons, or if pre-existing relativistic seed electrons are required. In some cases AGN can provide such a population of sub-GeV electrons. However, it is unclear how common this connection is. In this paper we present LOFAR 140 MHz and VLA L-band radio observations, as well as \textit{Chandra} data of \target{}, a merging galaxy cluster system hosting a pair of radio relics. A large patch of diffuse emission connects a bright radio galaxy with one of the relics, likely affecting the properties of the relic. We find that the most plausible explanation for the connection is that the merger shock wave has passed over an AGN lobe. The shock passing over this seed population of electrons has led to an increased brightness in the relic only in the region filled with seed electrons.   
\end{abstract}


\section{Introduction}
\label{sec:introduction}
Mergers of galaxy clusters are energetic astronomical events in which a fraction of the energy budget is dissipated into the acceleration of charged particles within the hot, magnetised intra-cluster medium \citep[ICM,][]{Brunetti2014}. This process results in the formation of Mpc-scale radio synchrotron emission that is broadly split into two categories: giant radio halos and relics \citep[also known as radio shocks; see][for a review]{VanWeeren2019}. Giant radio halos are generally characterised by low surface brightness, unpolarised emission spanning the cluster centre, approximately tracing the thermal bremsstrahlung X-ray radiation emitted by the baryonic mass distribution of the ICM. Giant radio halos are thought to be generated by turbulent (re-)acceleration of relativistic electrons injected into the ICM by mergers. \citep[e.g.][]{Brunetti2001,Petrosian2001}. 
Radio relics are observed primarily as elongated arcs of polarised emission in cluster outskirts. They are thought to trace merger shock waves as they propagate through the ICM \citep[e.g.][]{Ensslin1998,Roettiger1999,VanWeeren2010}. Observations of discontinuities in the thermodynamical properties of the ICM (density, pressure, temperature), coinciding with relics, support this scenario \citep[e.g.][]{Finoguenov2010,Bourdin2013,Akamatsu2017,Urdampilleta2018}. Diffusive Shock Acceleration \citep[DSA, see][for a review]{Blandford1987} is the most likely mechanism driving the particle acceleration at shock fronts. It can explain many properties of relics, such as their morphologies, radio fluxes, polarisation properties and the power-law energy distribution of cosmic rays \citep{Hoeft2007, Kang2012, Bruggen2020}. Observations of some cluster relics suggest that they can be explained by diffusive shock acceleration of electrons from the thermal pool (standard DSA), as in e.g. Abell 2249 \citep{Locatelli2020}, El Gordo and Abell 521  \citep{Giacintucci2008, Botteon2016, Botteon2020}. However, in general this is not the case. The Mach numbers of shocks generated by galaxy cluster mergers are typically low ($\mach \lesssim$ 5), so the acceleration efficiencies required to reproduce the luminosity of radio relics are too large to be explained by standard DSA \citep[e.g.][]{Kang2012, Brunetti2014, Vazza2014, Eckert2016, Botteon2020}. Moreover, in the DSA scenario gamma rays should be produced by collisions between shock-accelerated protons and thermal ICM protons. However, calculations suggest that an abnormally high electron-proton ratio at cluster shocks would be required to reconcile relics with the non-detection of gamma rays from clusters \citep{Ackermann2010, Vazza2014}.

The re-acceleration of a pre-existing population of sub-GeV electrons could mitigate the requirement of high acceleration efficiency at weak cluster shocks \citep[e.g.][]{Markevitch2005, Kang2012, Pinzke2013, Kang2014}. In a few clusters there is evidence of relativistic electrons from the tail of a radio galaxy being re-accelerated by a passing merger shockwave, as in Abell 3411-3412 \citep{VanWeeren2017}, PLCKG287.0+32.9 \citep{Bonafede2014} and CIZA J2242.8+5301 \citep{Gennaro2018}, for example. However a connection between a radio galaxy tail and a relic is not observed in all galaxy clusters hosting a relic, so it is unclear if this can fully solve the problem. Furthermore, re-acceleration does not solve the non-detection of gamma rays at cluster shocks. Unless galaxy jets and lobes are lepton-dominated, protons are also expected to be re-accelerated and therefore produce gamma rays via hadronic collisions \citep[][]{Vazza2014}.  

The advent of low frequency radio observatories has allowed new insights into the mechanisms producing large-scale emission in galaxy clusters and led to the discovery of new steep-spectrum sources, essentially invisible to even the most sensitive higher frequency observatories \citep[e.g.][]{DeGasperin2017, Mandal2020}. These fossil plasma sources might be revealing an additional population of seed relativistic electrons that are available to be re-accelerated by either turbulence or a passing shock wave. As more of these sources are detected, and we are better able to understand the prevalence of fossil plasma pools in clusters, their relevance to other cluster radio emission should become more apparent.

Approximately one third of clusters currently known to contain a radio relic host a pair of diametrically-opposed relics, so-called double radio relics \citep{DeGasperin2014, VanWeeren2019}. From simulations such clusters should be undergoing clean binary mergers on, or close to, the plane of the sky \citep[e.g.][]{VanWeeren2011}. Such constraints on the viewing angle minimises the influence of projection effects on analysis. The simple merger geometry makes clusters hosting double radio relics ideal for studying not only relativistic particle (re-)acceleration, but also gas and dark matter distributions \citep[e.g.][]{Golovich2019}.  

In this paper we present follow-up radio observations of \target{}, a low-mass, double radio relic merging galaxy cluster system, with the Karl G. Jansky Very Large Array (VLA), the Low Frequency Array (LOFAR) and X-ray observations with \textit{Chandra}.

\subsection{PSZ2 G096.88+24.18}

\target{} \citep[also known as ZwCL1856.8+6616][$z = 0.3$]{Zwicky1961} was detected through the Sunyaev-Zeldovich (SZ) effect by the \textit{Planck} satellite \citep[][]{Planck2011}. A total SZ mass of $M_{500}$\footnotemark $=(4.7\pm0.3)\times10^{14}\Msun$ was reported in the second \textit{Planck} data release \citep[][]{Planck2016}.

\footnotetext{$M_{500(200)}$ is the mass enclosed within the radius $r_{500(200)}$, at which the mean density of the cluster is 500(200) times the critical density of the Universe at the cluster redshift.}

\citet{DeGasperin2014} discovered a pair of radio relics on the northern and southern edges of \target{} at 1.4 GHz with Westerbork Synthesis Radio Telescope (WSRT). They also reported a low significance detection of a giant radio halo. Recently \citet{Finner2020} investigated the merger scenario of \target{} by combining weak gravitational lensing analysis using \textit{Subaru}, archival X-ray \textit{XMM Newton} data and the radio intensity images presented fully in this paper. They found that the radio relics, mass distribution from weak lensing and X-ray morphologies of \target{} are all aligned along the same axis. Combined with the spectroscopic results of \citet{Golovich2019a}, who found a single-peak redshift distribution of galaxies associated with \target{}, the merger is likely a head-on collision on the plane of the sky. \citet{Finner2020} also found that the time since collision is $0.7^{+0.3}_{-0.1}$ Gyr and the mass ratio of the merging subclusters is 1:1, with a total mass of $M_{200}$\footnotemark[\value{footnote}] $= 2.7^{+1.1}_{-1.5}\times 10^{14} \Msun $.
The two mass estimates of \target{} do not agree. However, cluster masses derived from the SZ effect are known to be significantly larger than those derived from weak lensing, primarily due to departures from hydrostatic equilibrium, to which SZ measurements are sensitive \citep[see][and references therein]{vonderLinden2014}.

This paper is organised as follows. In Section~\ref{sec:observations} we detail the LOFAR, VLA and \textit{Chandra} observations and their data products. We present our results for \target{} in Section~\ref{sec:results}. The discussion and conclusions of our results are in Sections~\ref{sec:discussion} and \ref{sec:conclusions}.
In this paper we assume a flat $\Lambda$CDM cosmology, with $H_0 = 70$ \kmsmpc{} and $\Omega_{M}$ = 0.3. At the redshift of \target{} ($z = 0.3$), 1\arcsec{} corresponds to a linear scale of 4.45 kpc.

\section{Observations \& Data Analysis}
\label{sec:observations}
In this paper we present LOFAR HBA 140 MHz and VLA 1.5 GHz observations of \target{}. Details of the observations can be found in Table~\ref{tab:Radio_obs}. Additionally we present \textit{Chandra} X-ray data of \target{}. In the following sections we summarise the main data calibration and imaging steps.

\subsection{LOFAR}
\label{sec:LOFAR obs}
\begin{table*}
 \centering
 \caption{Radio Observations}
 \label{tab:Radio_obs}
 %\resizebox{\columnwidth}{!}{%
 \begin{tabular}{lcccccccc}
  \hline
  Telescope & Project ID & Frequency & Array Mode/Configuration & Observation Date & Integration Time & On-Source Time & Primary Calibrator\\
  & & [GHz]  & & & [s] & [hr] & \\
  \hline
  LOFAR & LC9\_036 & 0.120 - 0.187 & HBA\_DUAL\_INNER & 2018 Aug 19 & 1 & 8 & 3C295\\
  VLA & 15A-056 & 1 - 2 & C & 2016 Feb 01 & 5 & 3.5 & 3C286\\
  VLA & 15A-056 & 1 - 2 & CnB-B & 2015 Feb 05 & 3 & 2.5 & 3C286\\
  VLA & 15A-056 & 1 - 2 & CnB-B & 2015 Feb 02-03 & 3 & 2.5 & 3C147\\

  \hline
 \end{tabular}
\end{table*}

\target{} was observed for 8 hrs on 2018 August 19 with LOFAR High Band Antenna (HBA) stations in HBA\_DUAL\_INNER mode as part of the project LC9\_036. The total 120 -- 187 MHz bandwidth was divided into 243 sub-bands. Prior to the target observation the calibrator 3C295 was observed for 10 minutes. Radio Frequency Interference (RFI) was removed using \textsc{aoflagger} \citep{Offringa2012}, as is done for all datasets before being archived. The standard LOFAR processing pipeline \textsc{prefactor3} \citep{DeGasperin2019} was used to calculate direction-independent instrumental and ionospheric effects, including the station-based clock offsets, polarisation offsets, bandpass and ionospheric rotation measure. The calibrator flux density scale was set according to \citet{Scaife2012}.

Earth's ionosphere has a greater effect on observations at low radio frequencies. LOFAR therefore requires advanced calibration techniques to adjust for direction-dependent effects caused by the ionosphere and additionally, imperfect beam models. Following the direction-dependent calibration procedure detailed in \citet{VanWeeren2016}\footnote{\url{https://github.com/lofar-astron/factor}}, the field around the target was split into roughly 60 facets, each containing a bright ($>0.3$~Jy) calibrator source. Self-calibration was performed on the calibrator source of the target facet, all those surrounding the target facet and any additional facets containing such a source with flux density of $\geq$1 Jy. The calibration solutions obtained for each calibrator were then applied to the whole facet. After self-calibration we achieve a typical resolution of $\sim 5$\arcsec{}, RMS noise of $\sim 0.11 $ \mjybeam{} and dynamic range of $\sim 5000$. The absolute flux scale is consistent, within errors, with the LOFAR Data Release 2 images (Shimwell et al., in prep.).

\subsection{VLA}
\label{sec:VLA obs}

Three L-Band (1-2 GHz) observations of \target{} were completed by VLA on separate days as part of project 15A-056, with a total on-source time of 8.5 hours. For each observation the full bandwidth was split into 16 spectral windows, each comprised of 64 channels of width 1 MHz. Details of the observations are given in Table~\ref{tab:Radio_obs}. All polarisation products were recorded.

The data were calibrated using \textsc{casa} version 5.3 \citep{McMullin2007}. Prior to calibration the setup scan, shadowed antennas, first and last integrations of each target scan and any zero amplitude data were flagged. The data were Hanning-smoothed and \textsc{aoflagger} was used to remove RFI from observations of the calibrator sources. The data were then calibrated for the gain curve, efficiency, antenna position offsets, global delays, bandpass and local gains using the primary and phase calibrators. The data were calibrated for cross-hand delays and polarisation angle using a calibrator with known polarisation fraction and angle and for instrumental polarisation using a calibrator with no, or very little, polarised emission. A list of the calibrators used for the three can be found in Table~\ref{tab:VLA_cals} The flux density was scaled to the primary calibrator models from \citet{Perley2017} and all calibrator solutions were applied to the target field. RFI in the target observations was removed using \textsc{aoflagger}. After averaging the data to 10 seconds the solutions for the target field were refined with several rounds of  phase and amplitude self-calibration. Once calibrated, the individual datasets were combined, resulting in a typical dynamic range of $\sim 900$.

Sources outside the VLA primary beam were subtracted from the $uv$-data to avoid contamination through the VLA beam sidelobes. Stokes I, Q, U and V images were obtained with \textsc{wsclean} \citep{Offringa2014} using various weightings and taperings (see Table~\ref{tab:Image_props}).

\begin{table*}
 \centering
 \caption{Radio Image Properties}
 \label{tab:Image_props}
 \begin{tabular}{lccccc}
  \hline
  Telescope & Beam & RMS Noise & Weighting & $uv$-Tapering & Discrete Sources Removed\\
   & [\arcsec{}$\times$\arcsec{}] & [\mujybeam{}] & & [\arcsec{}] & \\
  \hline
  LOFAR & \beamtable{13}{7} & 127 & Briggs, robust = 0 & - & No \\
  LOFAR & \beamtable{20}{17} & 179 & Briggs, robust = 0.3 & - & Yes/No \\
  LOFAR & \beamtable{15}{9} & 135 & Briggs, robust = 0.1 & - & Yes \\
  LOFAR & \beamtable{27}{24} & 213 & Briggs, robust = 0.4 & - & Yes \\
  VLA & \beamtable{10}{9} & 8 & Briggs, robust = 0 & 10 & No \\
  VLA & \beamtable{25}{25} & 32 & Briggs, robust = 0 & 30 & Yes \\
  \hline
 \end{tabular}
\end{table*}

\subsection{Radio Data Products}
\label{sec:Radio data prods}

All radio continuum images were produced using \textsc{wsclean}, with different weighting schemes to produce images with different resolutions. All images have a central frequency of 140 MHz and 1.5 GHz for LOFAR and VLA respectively. A list of all radio continuum images included in this paper, along with their most important imaging parameters and properties, can be found in Table~\ref{tab:Image_props}. All images were corrected for primary beam attenuation.

\begin{figure*}
    \centering
    \includegraphics[width=\columnwidth]{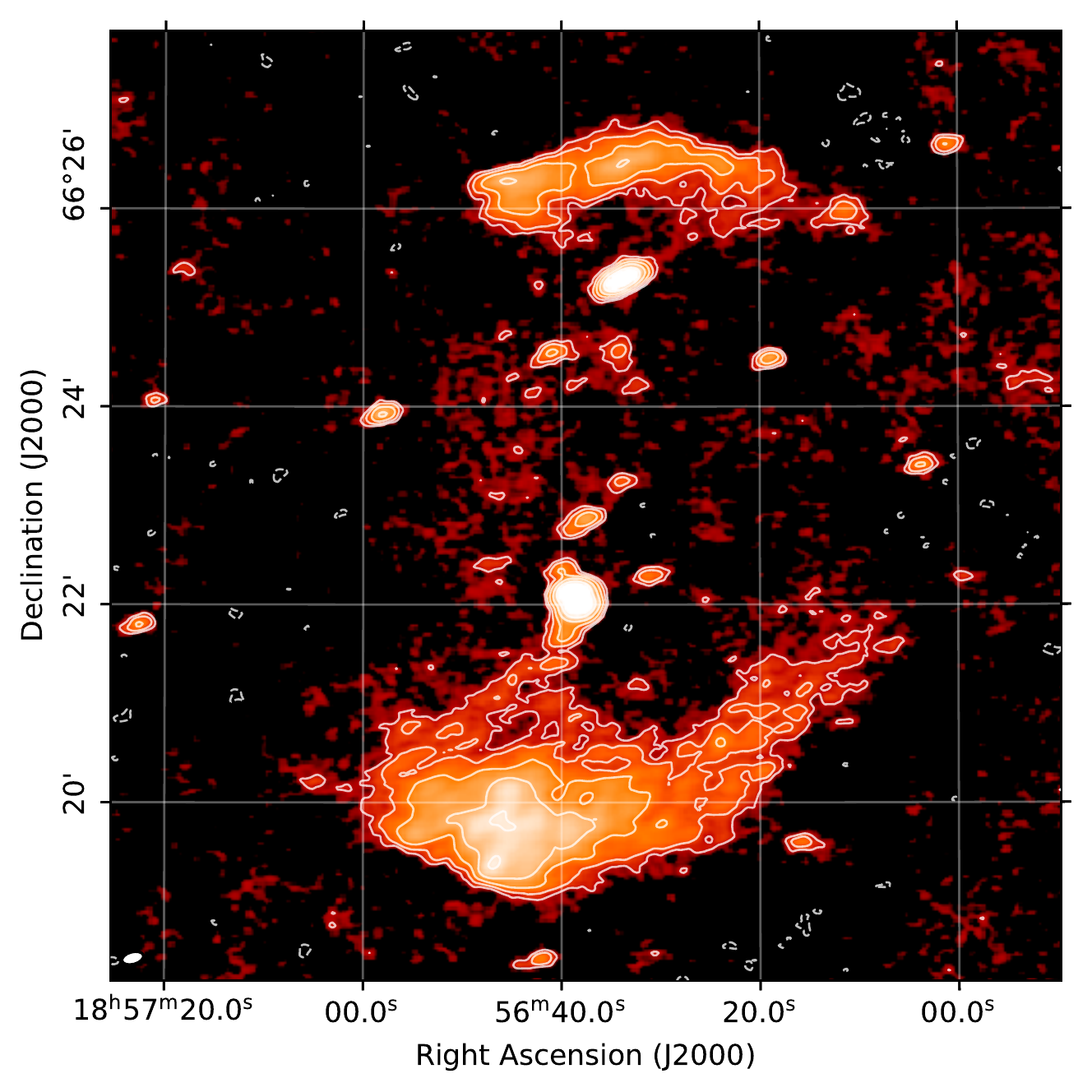}
    \includegraphics[width=\columnwidth]{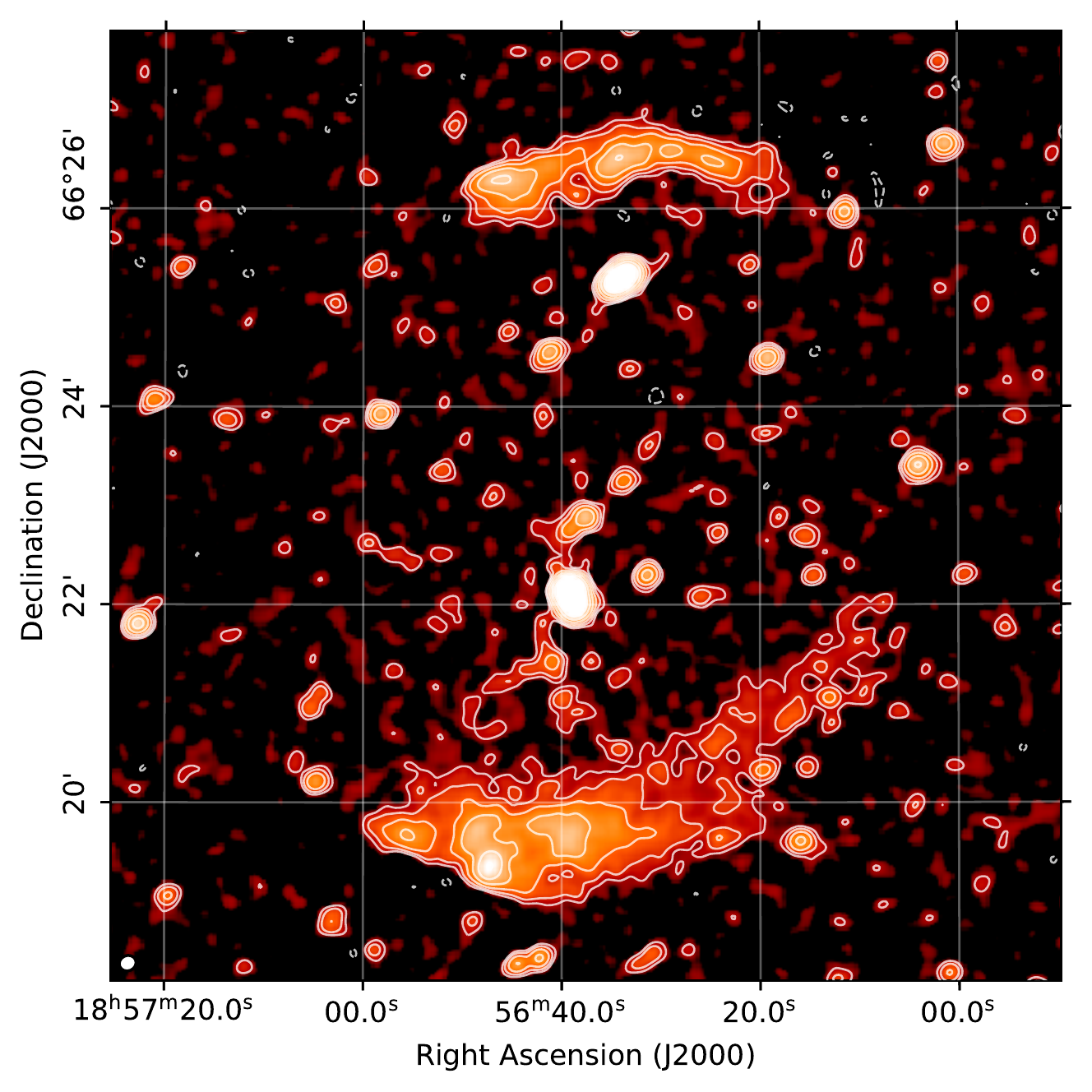}
    \caption{Radio intensity images of \target{}. Left: LOFAR 140 MHz (resolution: \beam{13}{7}, RMS noise: 127 \mujybeam{}. Right: VLA 1.5 GHz (resolution: \beam{10}{9}, RMS noise: 8 \mujybeam{}). Contour levels for both images are set at $3\sigma_{\textrm{rms}} \times$ [1,2,4...]. Dashed contours at $-3\sigma_{\textrm{rms}}$. Beam shapes are drawn in the bottom left corner of each image.}
    \label{fig:Radio_images}
\end{figure*}

Radio intensity images of \target{} from both LOFAR and VLA are shown in Fig.~\ref{fig:Radio_images}.

The error in measured flux density is calculated using

\begin{equation}
\label{eq:flux err}
\sigma_{S} = \sqrt{\left(\sigma_{\rm{rms}}\sqrt{n_{\rm{beams}}}\right)^2 + \left(\sigma_{\rm{cal}}\right)^2} ,
\end{equation}
where $\sigma_{\rm{rms}}$ is the RMS noise, $n_{\rm{beams}}$ the number of beams within the region of interest and $\sigma_{\rm{cal}}$ the flux error from calibration. We assumed a 15\% flux calibration error for LOFAR \citep[][]{Shimwell2016} and 5\% for VLA \citep[][]{Perley2017}.

To remove the contribution of discrete sources embedded within the diffuse emission we subtracted them from the $uv$-data. To achieve this the data were imaged at high resolution, with an inner $uv$-cut of 5.15 \klambda{}. At z = 0.3 this corresponds to excluding scales above a linear size of 178 kpc. The $uv$-cut was chosen to exclude most of the diffuse cluster emission. \textsc{pybdsf} \citep{Mohan2015} was used to detect $5\sigma$ point sources in the image and the results were checked by eye. We ignored detections within the radio relics as they likely correspond to the brightest regions of diffuse emission, while the rest of the relic is mostly resolved out in the high resolution image. All remaining point sources were subtracted from the $uv$-data using the clean components created by \textsc{wsclean}.

\subsubsection{Spectral Index}
\label{sec:Spectral Index}

To account for the different $uv$-coverages of LOFAR and VLA we imaged both datasets, with discrete sources subtracted, within the same $uv$-range and tapered to approximately the same beam size, using Briggs weighting, robust=-0.5 \citep[][]{Briggs1995}. The images were subsequently convolved to the same beam (medium resolution: \beam{10}{10}, low resolution: \beam{25}{25}).

\begin{figure}
    \centering
    \includegraphics[width=\columnwidth]{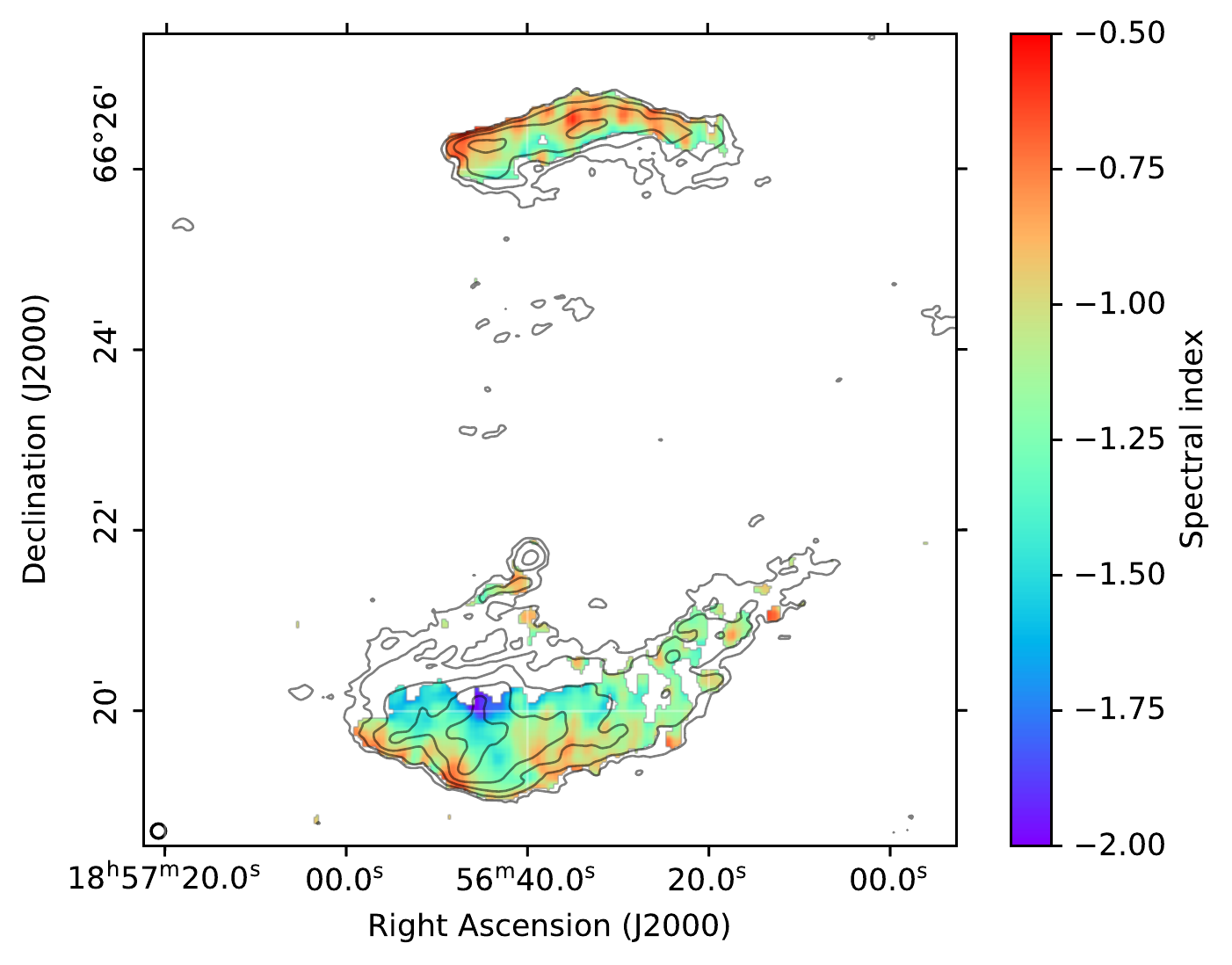}
    \includegraphics[width=\columnwidth]{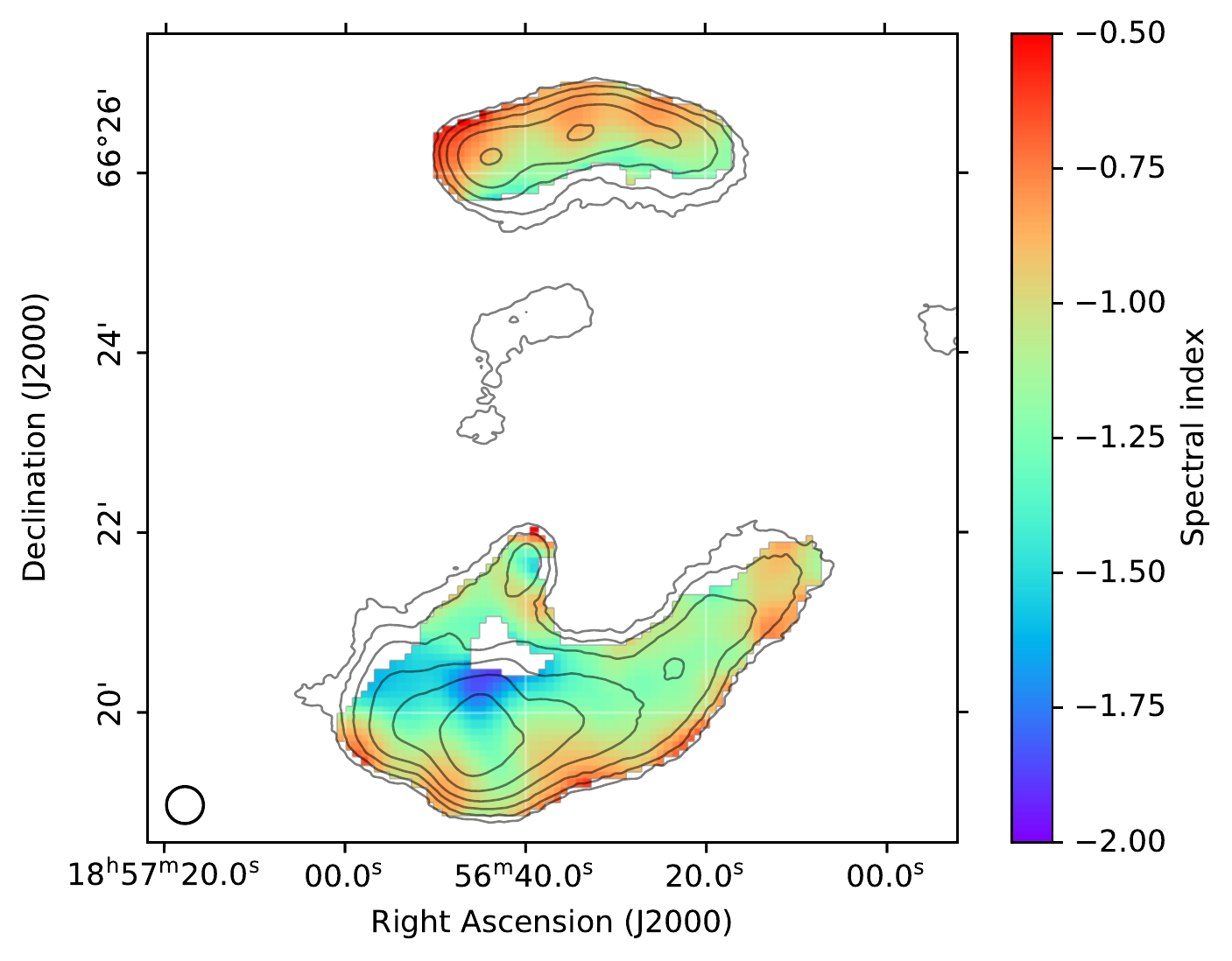}
    \caption{Spectral index maps created between 140 MHz and 1.5 GHz. Top: High resolution (\beam{10}{10}). Bottom: Low resolution (\beam{25}{25}). Corresponding error maps are shown in Fig.~\ref{fig:Spidxerrmaps}.}
    \label{fig:Spidxmaps}
\end{figure}

To produce maps of the spectral index across \target{} we computed the spectral index $\alpha^{140}_{1500}$ for each pixel, excluding any pixels with a flux density $<3 \sigma_{\rm{rms}}$ in either image. Throughout the paper we calculate the spectral index error as
\begin{equation}
\label{eq:spidx err}
\sigma_{\alpha} = \frac{1}{\ln{\frac{\nu_1}{\nu_2}}} \sqrt{\left(\frac{\sigma_{S,1}}{S_1}\right)^2 + \left(\frac{\sigma_{S,2}}{S_2}\right)^2},   
\end{equation}
where $\nu_1$ and $\nu_2$ are the characteristic frequencies of the images, $S_1$ and $S_2$ are the flux densities and $\sigma_{S,1}$ and $\sigma_{S,2}$ the respective flux density errors calculated from Equation~\ref{eq:flux err}.
The spectral index maps produced are shown in Fig.~\ref{fig:Spidxmaps}.

\subsubsection{Polarisation}
\label{sec:Polarisation}

All four correlation products were recorded by VLA, so we were able to calculate all Stokes parameters I,Q,U and V.
Assuming negligible circular polarisation, fractional polarisation is given by

\begin{equation}
\label{eq:pol frac}
P_{\rm{frac}} = \frac{S_{\rm{Pol}}}{S_I},    
\end{equation}
where $S_I$ is the total intensity and $S_{\rm{Pol}}$ the true polarised intensity.
At low signal-to-noise the noise can bias the measured polarisation intensity to higher values \citep[][]{Rice1945,Simmons1985}. To account for this we calculate the true polarisation intensity from the measured intensity using the results of \citet{Wardle1974}:

\begin{equation}
\label{eq:Rice_bias}
S_{\rm{Pol}} \sim S^{\prime}_{\rm{Pol}}\sqrt{1-\left(\frac{\sigma^{\prime}_{\rm{Pol}}}{S^{\prime}_{\rm{Pol}}}\right)^{2}}, 
\end{equation}
where $\sigma^{\prime}_{\rm{Pol}}$ is the measured RMS noise in polarisation intensity and $S^{\prime}_{\rm{Pol}}$ the measured polarisation intensity, which is calculated from the measured linear polarisation intensities $S_Q$ and $S_U$ (Stokes parameters Q and U) using

\begin{equation}
\label{eq:Meas_pol_int}
S^{\prime}_{\rm{Pol}} = \sqrt{\left(S_Q\right)^2 + \left(S_U\right)^2}.
\end{equation}
The polarisation angle is then calculated using 

\begin{equation}
\label{eq:pol ang}
\Psi_{\rm{pol}} = \frac{1}{2} \arctan\left(\frac{S_U}{S_Q}\right).
\end{equation}
Ionised material, in the presence of a magnetic field, along the line of sight between the observer and the target will rotate the intrinsic polarisation angle by an amount

\begin{equation}
\label{eq:Far rot}
\Delta\Psi = \lambda^2\textrm{RM},
\end{equation}
where $\lambda$ is the observation wavelength and RM is the rotation measure. From the map of galactic Faraday rotation presented in \citet{Oppermann2015}, we find that our Galaxy accounts for an RM $= -4.5$ rad m$^{-2}$ in the direction of \target{}, corresponding to a rotation of the polarisation angle $\Delta\Psi = -10 \deg$ at 1.5 GHz. We therefore rotate all polarisation vectors by this amount, to show only the intrinsic polarisation angle.

\begin{figure}
    \centering
    \includegraphics[width=\columnwidth]{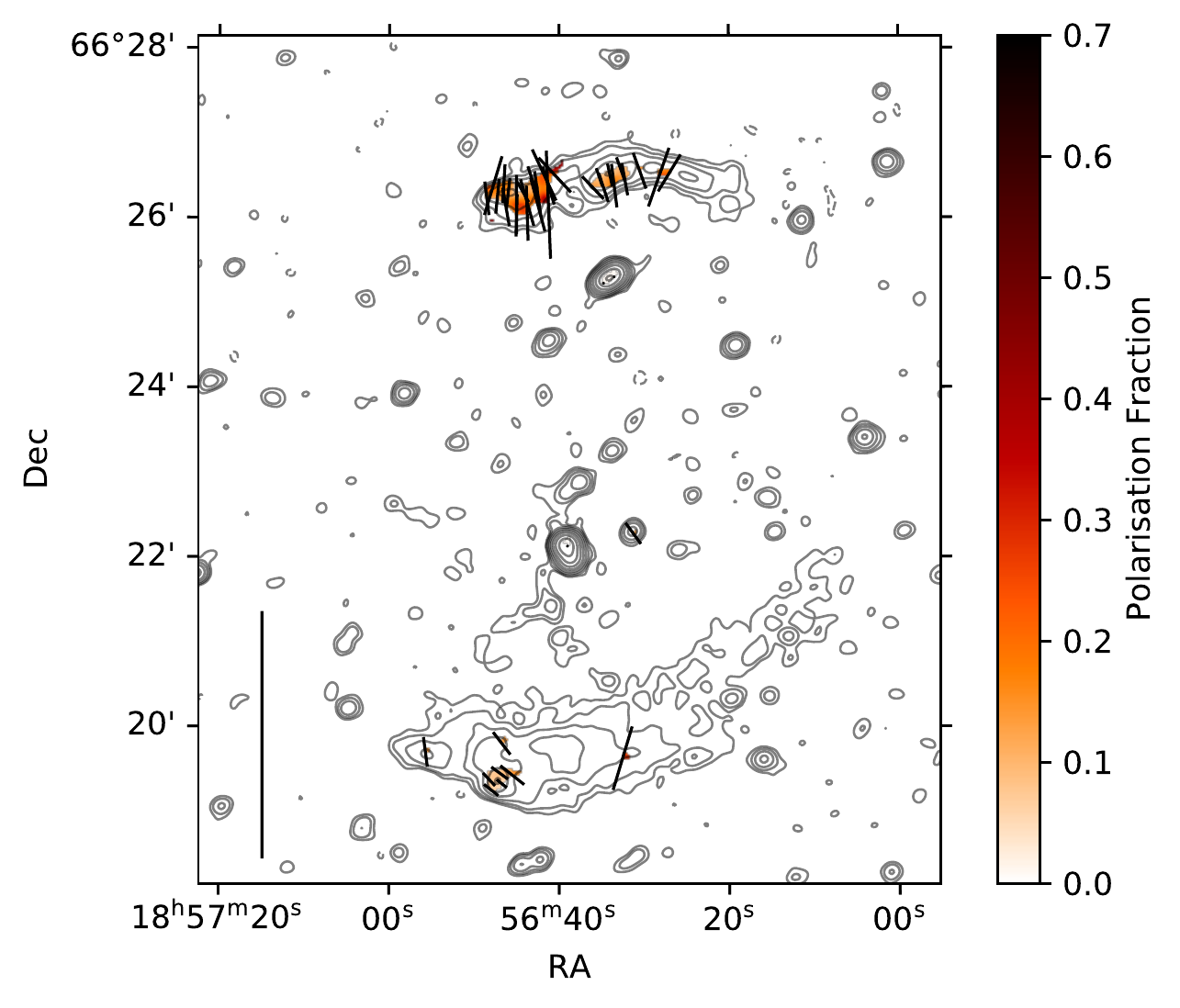}
    \caption{Linear polarisation fraction image. E-field vectors are plotted in black, with amplitude proportional to the polarisation fraction. A reference vector, corresponding to 100\% polarisation fraction, is plotted in the bottom left. Contours as in Fig.~\ref{fig:Radio_images} (right).}
    \label{fig:Polarisation Fraction}
\end{figure}

Fig.~\ref{fig:Polarisation Fraction} shows the polarisation vectors, corresponding to the electric field vectors, overlaid on an image of the fractional polarisation measured at 1.5 GHz by VLA.

\subsection{\textit{Chandra}}
\label{Chandra obs}
\begin{figure}
    \centering
    \includegraphics[width=\columnwidth]{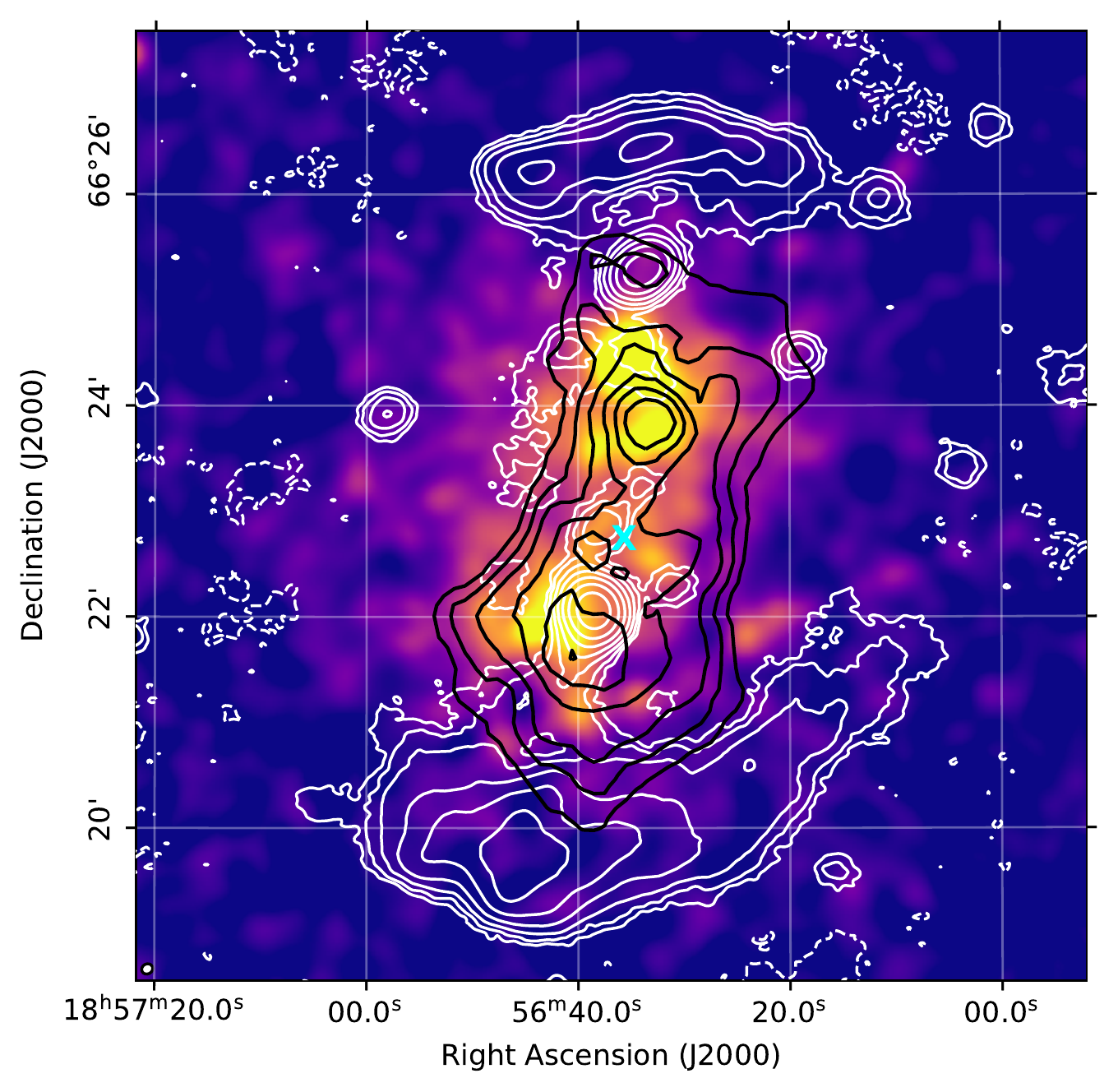}
    \caption{Exposure-corrected Chandra X-ray $1.0 - 4.0$ keV image. The image is smoothed with a gaussian kernel of width $\sigma = 7\arcsec{}$. Black contours show the mass distribution from \citet{Finner2020}. Levels at: $\sigma_{\rm{mass}} \times$ [2.1,2.4,2.7,3.1,3.4]. White contours from the LOFAR 140 MHz image (resolution: \beam{20}{17}, RMS noise: 179 \mujybeam{}). Contour levels at $3\sigma_{\rm{rms}} \times$ [1,2,4...]. Dashed contours at $-3\sigma_{\rm{rms}}$. The cyan cross denotes the cluster centre used in this work.}
    \label{fig:Radio_Mass_Xray}
\end{figure}

On 2018 June 03 \textit{Chandra} ACIS-I detector observed \target{} for 43 ks (ObsID 19752). The data were calibrated with \textsc{ciao} version 4.12 \citep{Fruscione2006} using \textsc{caldb} version 4.9.1. The event files were reprocessed using \textsc{chandra\_repro} to create new level 2 data products. The data were filtered to $0.5-7$ keV to exclude contamination at low energies and the high energy particle background. Periods with anomalously low count rates and flares were found and removed using \textsc{lc\_clean}. 42ks remained after filtering. Blank sky event files were used to model the instrumental background on the observation date. Point sources were detected using \textsc{wavdetect} with wavelet scales of 1, 2, 4, 8 and 16 pixels, visually inspected and excluded for analysis. The $1 - 4$ keV \textit{Chandra} exposure-corrected flux image is shown in Fig.~\ref{fig:Radio_Mass_Xray}, with radio and weak lensing mass distribution contours overlaid.

\section{Results}
\label{sec:results}

\subsection{Radio Relics}
\label{sec:radio relics res}

\begin{figure*}
    \centering
    \includegraphics[width=\textwidth]{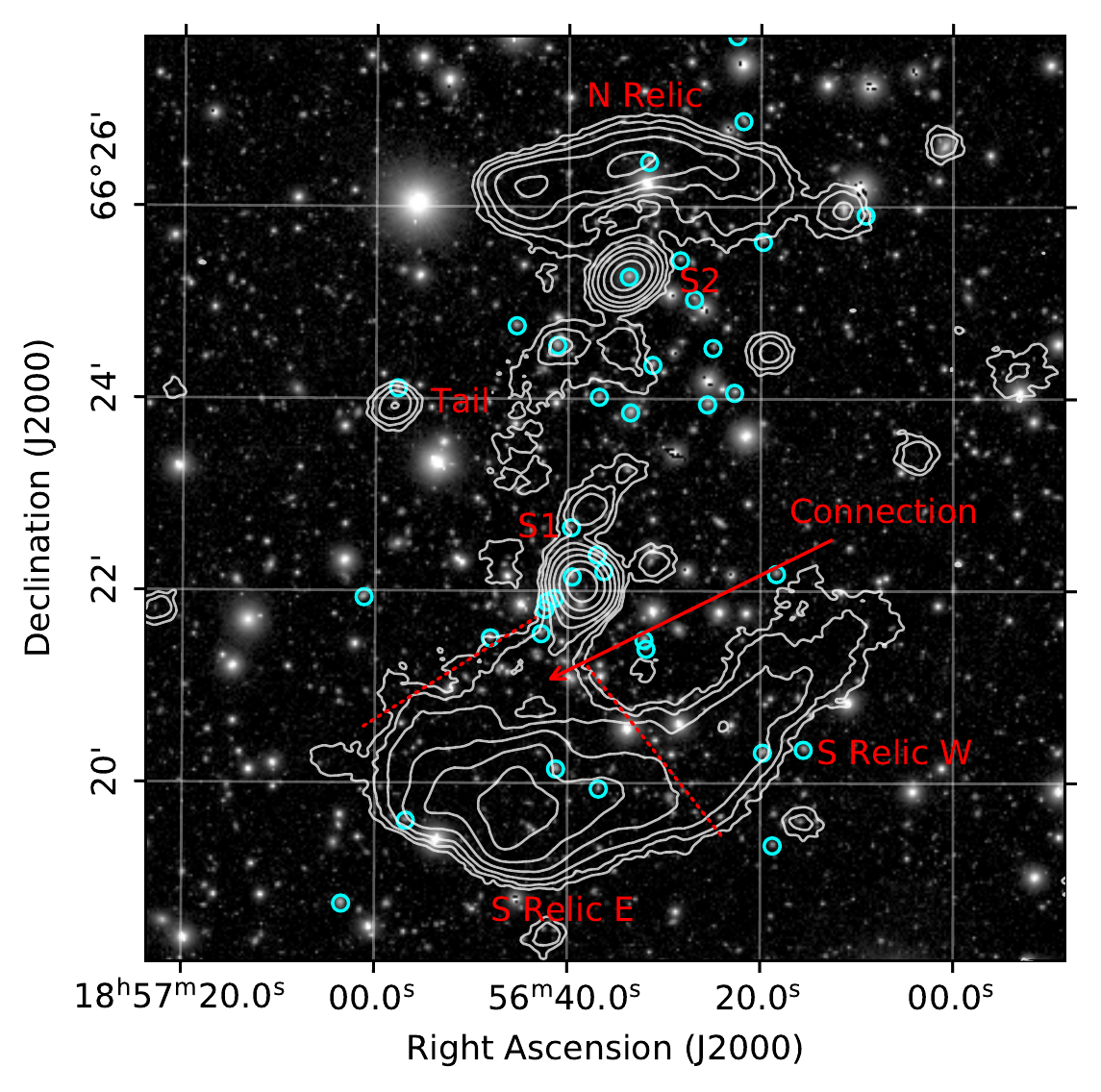}
    \caption{Subaru Suprime r-band image from \citet{Finner2020}. Overlaid LOFAR contours as in Fig.~\ref{fig:Radio_Mass_Xray}. Radio sources of particular relevance to this paper are labelled. Dashed lines are drawn to follow the Connection contours, extrapolated to the radio relic edge. Cyan circles show the spectroscopically-confirmed cluster members.}
    \label{fig:Radio optical labelled}
\end{figure*}

Fig.~\ref{fig:Radio optical labelled} shows LOFAR radio contours overlaid on an r-band Subaru image. Regions of particular relevance to this paper are labelled and spectroscopically-confirmed cluster-member galaxies are highlighted \citep[][]{Golovich2019a}. The image shows the presence of two arc-like structures on opposite sides of the cluster (N and S relics). The morphology, location at the cluster periphery and spectral index variation (Fig.~\ref{fig:Spidxmaps}) of the arc-like radio structures in \target{}, a merging galaxy cluster, all strongly suggest that we are observing a pair of radio relics. If we define the cluster centre as the peak of X-ray emission the outer edges (shock fronts) of the northern and southern relics lie at a distance of $\sim$ 560 kpc and 1.3 Mpc respectively. However \target{} is a galaxy cluster comprised of two merging subclusters separated by $\sim$ 600kpc \citep{Finner2020}. The X-ray peak is therefore located at the centre of the brightest subcluster. If instead we take the cluster centre to be at the midpoint between the two mass peaks (cyan cross in Fig.~\ref{fig:Radio_Mass_Xray}), the shock fronts of the northern and southern relics lie at a distance of $\sim$ 1.1 and 1.0 Mpc from the cluster centre. We choose the latter as our definition of the cluster centre. 

At 140 MHz the radio relics have largest linear sizes (LLS) of $\sim$ 0.9 and 1.5 Mpc and a flux ratio of 1:3.5 for the north and south relics respectively. Both radio relics have a non-uniform brightness along their major axes (Fig.~\ref{fig:Radio_images}). The average surface brightness is approximately $4$ times greater in the eastern side of the southern relic compared to the west. Higher resolution images (Fig.~\ref{fig:Radio_images}) of \target{} show filament-like substructures in the northern relic.

\begin{figure}
    \centering
    \includegraphics[width=\columnwidth]{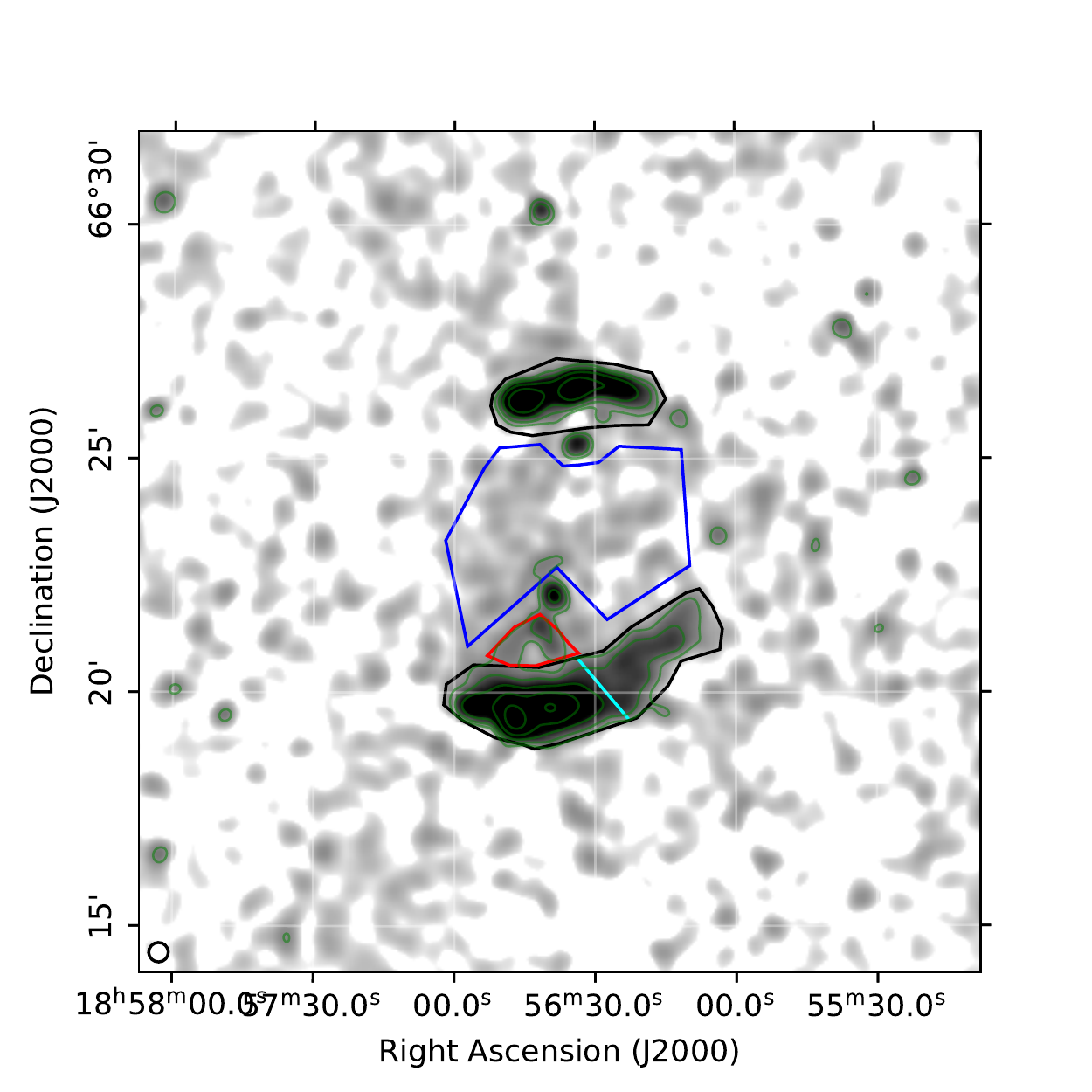}
    \caption{VLA intensity image (resolution: \beam{25}{25}, RMS noise: 32 \mujybeam{}). Contour levels at $3\sigma_{\rm{rms}} \times$ [1,2,4...]. Dashed contours at $-3\sigma_{\rm{rms}}$. Regions of extraction for the flux density measurements in Table~\ref{tab:Shock_props} in black (radio relics), red (Connection), cyan (line dividing east and west sides of the southern radio relic) and for the candidate radio halo in blue (see Section~\ref{sec:Halo}).}
    \label{fig:Extraction_Regions}
\end{figure}

The flux densities, spectral indices and Mach numbers of all regions of diffuse emission are listed in Table~\ref{tab:Shock_props}. Fig.~\ref{fig:Extraction_Regions} shows the regions used to calculate the flux densities, overlaid on a low resolution (\beam{17}{17}) VLA image.

\begin{table*}
 \centering
 \caption{Radio Diffuse Emission Properties}
 \label{tab:Shock_props}
 \begin{tabular}{lcccccc}
  \hline
   Source & $S_{\rm 140\,MHz}$ & $S_{\rm 1.5\,GHz}$ & $\alpha_{\rm{int}}$ & $\alpha_{\rm{inj}}$ & $\mach{}_{\rm{int}}$ & $\mach{}_{\rm{inj}}$\\ 
   & [mJy] & [mJy] & & \\
  \hline
  N Relic & $76 \pm 12$ & $7.8 \pm 0.4$ & $-0.95 \pm 0.07$ & $-0.87 \pm 0.07$ & - & $2.5 \pm 0.2$\\
  S Relic & $276 \pm 42$ & $16.5 \pm 0.9$ & $-1.17 \pm 0.07$ & $-0.97 \pm 0.07$ & $3.6 \pm 0.7$ & $2.3 \pm 0.2$ \\
  S Relic E & $234 \pm 35$ & $13.5 \pm 0.7$ & $-1.19 \pm 0.07$ & $-0.96 \pm 0.07$ & $3.4 \pm 0.6$ & $2.3 \pm 0.2$\\
  S Relic W & $38 \pm 6$ & $2.8 \pm 0.2$ & $-1.08 \pm 0.07$ & $-1.01 \pm 0.09$ & $5.1 \pm 2.2$ & $2.2 \pm 0.2$ \\
  Connection & $14 \pm 2$ & $0.7 \pm 0.1$ & $-1.22 \pm 0.08$ & - & - & -\\
  \hline
 \end{tabular}
\end{table*}

Fig.~\ref{fig:Spidxmaps} shows the spectral index maps at medium ($10$\arcsec{}) and low resolution ($25$\arcsec{}). Both the north and south relics exhibit spectral steepening from the shock edge towards the cluster centre, from about $\alpha = -0.6$ to $\alpha = -2.0$ in the case of the southern relic and $\alpha = -0.5$ to $\alpha = -1.5$ for the northern relic. A spectral steepening towards the cluster centre is expected from downstream synchrotron and Inverse Compton (IC) losses, as observed in many other clusters \citep[e.g.][]{Bonafede2012,VanWeeren2016a,Hoang2017,Gennaro2018}. Whilst the radio relics show a general steepening of the spectral index towards the cluster centre, the distribution in the southern relic is non-uniform. There are regions up to $\sim$ 40 \arcsec{} (170 kpc) downstream of the shock front with relatively flat ($\alpha > -1$) spectral indices (red in Fig.~\ref{fig:Spidxmaps}), as well as regions at the relic edge with relatively steep spectral index ($\alpha \sim -1.1$, green in Fig.~\ref{fig:Spidxmaps}).

In Fig.~\ref{fig:Shock_Spidx_One_Graph} we plot the integrated spectral index as a function of distance from the shock front in three different regions of the radio relics (blue, red and yellow). The short side of each region is one restoring beam width and the regions were oriented to follow the direction perpendicular to the shock front, moving downstream of both relics. Using the blue and red sets of regions we are able to compare the spectral behaviour of the east and west sides of the southern relic. In all three sets of regions (blue, red and yellow) the spectrum steepens towards the cluster centre. Up to the southern edge of the Connection (see Section~\ref{sec:Connection}), the east and west sides of the southern relic show similar steepening as a function of distance from the shock front.

Fig.~\ref{fig:Polarisation Fraction} shows the linear polarisation fraction image of \target{} at 1.5 GHz, with electric field vectors overlaid. In the northern radio relic there are significant regions of polarised emission, mostly corresponding to the brightest parts of the relic. The linear polarisation fraction ranges from $\sim$ 10\% - 60\%. The polarisation vectors, chosen here to show the electric field direction, are all approximately perpendicular to the shock front. This implies that the magnetic field is ordered and compressed along the shock front, as a consequence of shock passage \citep[][]{Ensslin1998}. 
The southern relic has polarised emission in only a few small regions and the polarisation fraction is much lower than in the northern relic, reaching a maximum of 20\%. In these small regions the electric field vectors do not lie perpendicular to the shock. These findings are in line with those of \citet{DeGasperin2014}, although obtained with a different dataset.

\subsection{Shock Mach Numbers}
\label{sec:Mach}

The strength of the shock, or Mach number $\mach{}$, affects the other observed properties of a shock front. DSA predicts that the slope of the spectral energy distribution at the shock front, $\alpha_{\rm{inj}}$, depends on the Mach number \citep[][]{Blandford1987}, with 

\begin{equation}
\label{eq:alpha mach}
    \alpha_{\rm{inj}} = \frac{1}{2} - \frac{\mach{}^{2} + 1}{\mach{}^{2} - 1}.
\end{equation}

If the injection spectral index can be measured accurately, this should provide the most reliable Mach number estimate (assuming DSA), as we expect the merger axis to be on the plane of the sky and therefore projection effects should be minimised. We do not have the frequency coverage to calculate $\alpha_{inj}$ by spectral ageing \citep[e.g.][]{Harwood2013, DeGasperin2015}. We instead choose to calculate the average spectral index along the shock front, using a region covering the entire shock front with a width of one beam.

An alternative method of estimating the Mach number of a shock is to use the integrated spectral index, $\alpha_{\rm{int}}$, as it reduces statistical uncertainties. By considering the shock as a simple planar shock, we can relate the integrated and injection spectral indices through 

\begin{equation}
\label{eq:alpha int}
    \alpha_{\rm{inj}} = \alpha_{\rm{int}} + \frac{1}{2}.
\end{equation}
\citep[][]{Kardashev1962}. However, such an approximation is likely not valid for spherically-expanding shocks \citep{Kang2015} or within a turbulent medium \citep{Dominguez-Fernandez2021}.
We present Mach numbers calculated by both methods in this paper in Table~\ref{tab:Shock_props}. 
It is also possible to independently measure the shock Mach number from X-ray surface brightness and/or temperature discontinuities. However, our \textit{Chandra} data is too shallow to detect any discontinuity and we are therefore unable to calculate Mach numbers in this way.

\subsection{Radio Relic Connection}
\label{sec:Connection}

\begin{figure}
\centering
\includegraphics[width=\columnwidth]{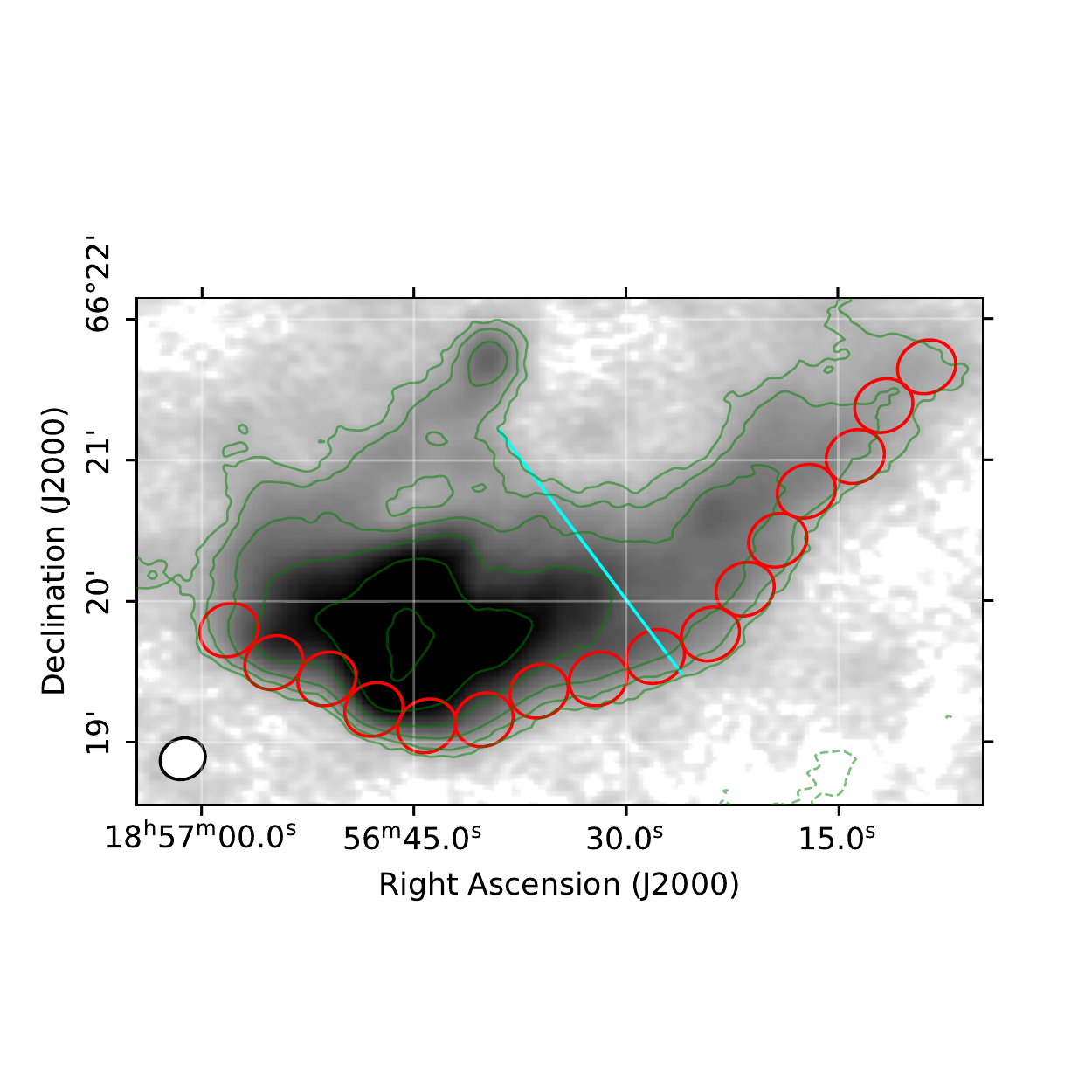}
\includegraphics[width=\columnwidth]{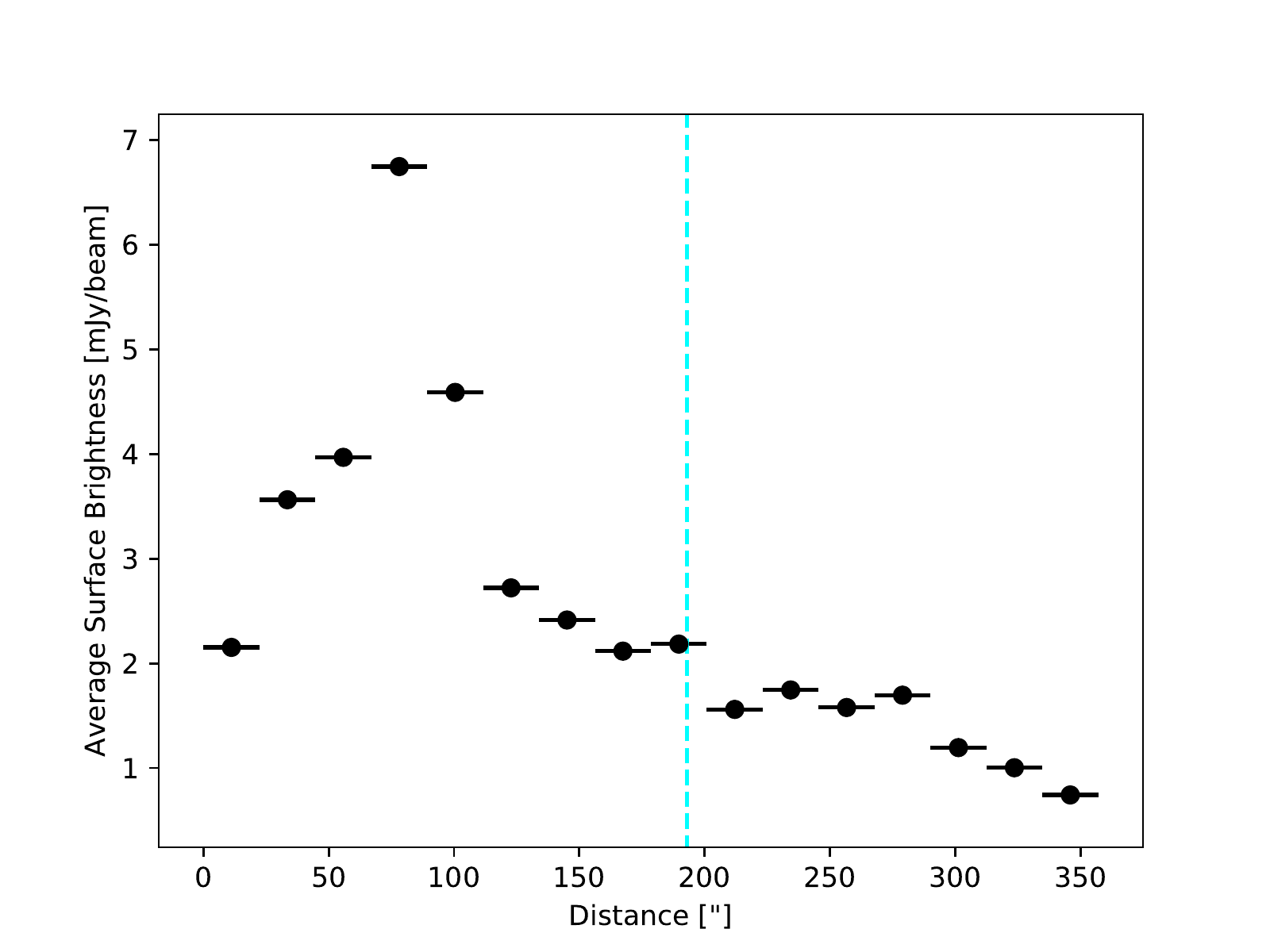}
\caption{Top: Regions in which we calculate the average surface brightness across the southern relic. Each region is the size of the synthesised beam. Bottom: The resulting surface brightness profile as a function of distance from the east edge of the shock front, along the axis of the shock front. The cyan line in both images shows the dividing line between the east and west sides of the relic, as in Fig.~\ref{fig:Radio optical labelled}.}
\label{fig:SRelic_SB_Profile}
\end{figure}

Between the southern radio relic and S1 there is a continuous patch of diffuse radio emission (Connection), clearly seen in the lower-frequency, LOFAR images (Fig.~\ref{fig:Radio optical labelled}). Due to its relatively steep spectrum and low surface brightness, the Connection is only visible in low resolution VLA images (Fig.~\ref{fig:Extraction_Regions}). There is a section of the Connection closest to the relic that is only visible in the LOFAR images. Fig.~\ref{fig:SRelic_SB_Profile} shows the variation of the average surface brightness along the shock front of the southern relic. The eastern side of the relic has an average flux density approximately 4 times greater than the western side and all individual regions have a greater average surface brightness than any on the western side. This surface brightness jump aligns perfectly with the Connection (Fig.~\ref{fig:Radio optical labelled}).

\begin{figure}
\centering
\includegraphics[width=\columnwidth]{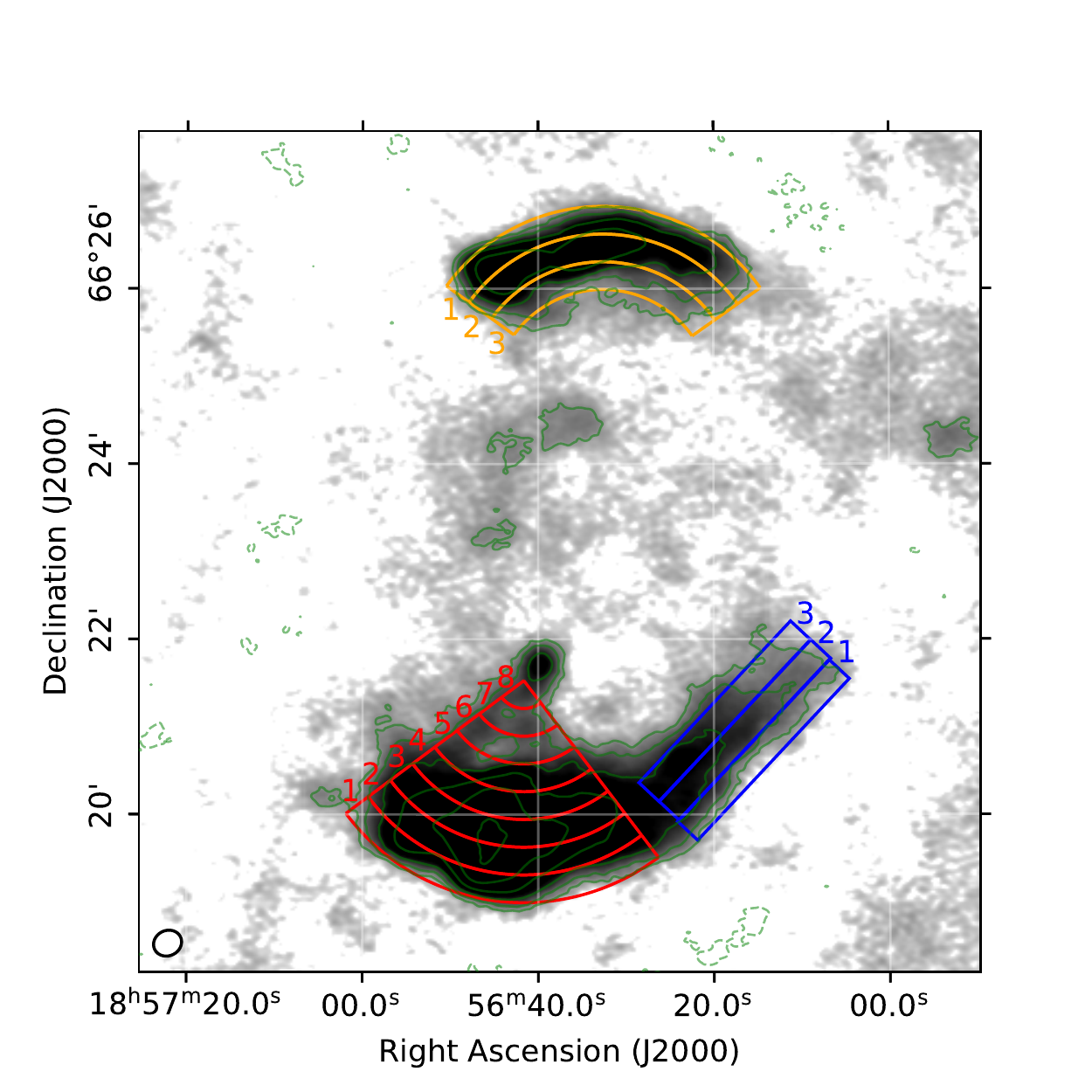}
\includegraphics[width=\columnwidth]{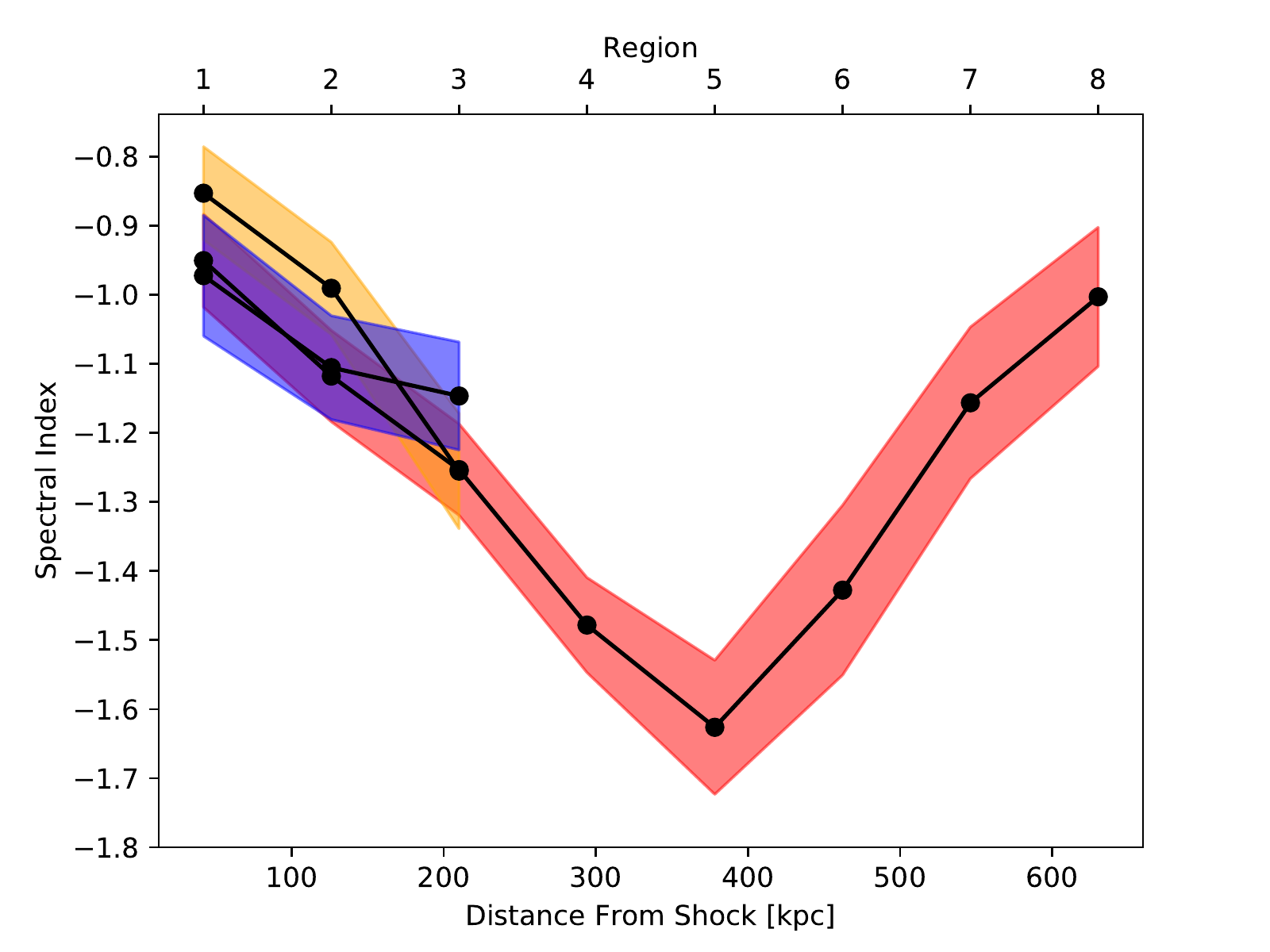}
\caption{Top: Regions where we extract the integrated spectral index. The short sides are exactly one beam width in length (19\arcsec). Bottom: The integrated spectral indices extracted from these regions as a function of distance from the shock front.}
\label{fig:Shock_Spidx_One_Graph}
\end{figure}

As there is a gap in Connection emission in the VLA images, the Connection is only visible in the low resolution spectral index map (Fig.~\ref{fig:Spidxmaps}, bottom panel) as two distinct bands of emission connecting S1 to the radio relic. The spectral index steepens along the eastern band, away from S1, from approximately $\alpha = -1.0$ to $\alpha = -1.3$. However, the western band varies between $\alpha = -0.8$ and $\alpha = -1.3$ with no clear spatial trend. The integrated spectral index along the Connection (Fig.~\ref{fig:Shock_Spidx_One_Graph}, red line) flattens from $\alpha = -1.6$ in region 5,$\sim 380$ kpc from the shock front, to $\alpha = -1.0$ in region 8, $\sim 620$ kpc from the shock front.

\section{Discussion}
\label{sec:discussion}

\subsection{Radio Relics}
\label{sec:radio relics dis}

We are unable to detect evidence of a shock front at the position of either radio relic in the X-ray data. Due to the low count statistics in cluster outskirts, the \textit{Chandra} data are likely not sensitive enough to detect a shock. \citet{Finner2020} also did not detect a shock with a 12 ks \textit{XMM-Newton} observation. The low electron density at the outskirts of galaxy clusters, where radio relics are typically located, makes detection of shocks with X-ray observations challenging. To date, a spatially-coincident X-ray shock has been discovered for only 20 relics \citep[][]{VanWeeren2019}. Nonetheless, the relic-shock connection is well-established \citep[see][for reviews]{Brunetti2014,VanWeeren2019}. In the specific case of \target{}, we expect the radio relics to be associated with merger shocks for a number of reasons, despite the non-detection in X-rays. From the spectral index distribution (Figs~\ref{fig:Spidxmaps} and \ref{fig:Shock_Spidx_One_Graph}), we see that the spectral index is flattest at the outer edges of both relics, furthest from the cluster centre and that the spectral index subsequently steepens towards the cluster centre. This spectral behaviour is exactly as expected for shock acceleration. The shock front, i.e. the shock edge furthest from the cluster centre, is the site of cosmic ray electron (re-)acceleration. After shock passage, cosmic ray electrons radiate energy away in the form of Inverse-Compton and synchrotron emission. In the absence of another acceleration mechanism, the spectral index of the region downstream of the shock front should therefore be steeper, due to these energy losses. 
The arc-like morphology of the radio relics in \target{} is similar to relics found in other clusters \citep[][]{VanWeeren2019} and to those found in simulations of merger shockwaves \citep[e.g.][]{Skillman2011,Nuza2017,Wittor2017}. The longest linear sizes and fluxes of the relics (see Sec.~\ref{sec:radio relics res}) are in line with the radio relic scaling relations of \citet{DeGasperin2014}, though it should be noted that the relics in \target{} were included in the sample used to create the scaling relations. Additionally, the location of the relics ($\gtrsim1$ Mpc from the cluster centre; see Sec~\ref{sec:radio relics res}) is in line with the results of \citet{Vazza2012} that, due to the higher kinetic energy dissipated, relics should be preferentially located in cluster outskirts. This also matches with observational findings.
Shock waves compress and align the magnetic field along the shock, resulting in high polarisation fractions and electric field vectors that lie perpendicular to the shock front \citep[][]{Brunetti2014}. The polarisation properties of the northern relic follow these expectations, but those of the southern relic do not (Fig~\ref{fig:Polarisation Fraction}). Cluster mergers are expected to launch a pair of shocks in opposite directions along the merger axis. However, favourable viewing conditions, with a merger axis on or almost on the plane of the sky, are required for both shocks to be observable in radio observations \citep[][]{VanWeeren2011}. From spectroscopic observations of \target{}, the merger is expected to be on the plane of the sky and we may therefore expect \target{} to host two relics \citep[][]{Golovich2019a}. Additionally, weak lensing analysis of \target reveals that the merger axis lies along the line between the north and south radio relics \citep[][]{Finner2020}. Given these pieces of evidence, we expect that the relics in \target{} trace shock waves launched into the ICM by a merger.

The shock wave should compress the magnetic fields within the cluster and align them approximately parallel to the shock front \citep[][and references therein]{VanWeeren2010,Brunetti2014}. This implies that the electric field vectors should lie perpendicular to the shock front and that a significant fraction of the detected radio emission should be polarised. Radio relics typically have polarisation fractions $\geq 10\%$ at GHz frequencies \citep[see][and references therein]{Wittor2019} but there are cases where the polarisation fraction can reach up to 70\% \citep[e.g.][]{Loi2017}.
We observe strongly polarised emission in the northern relic (Fig.~\ref{fig:Polarisation Fraction}), with all electric field vectors lying approximately perpendicular to the shock front, as expected. 
Conversely the southern relic shows very little polarised emission. The few polarised regions in the southern relic have very low polarisation fractions ($\leq 10\%$) and show an offset between the electric field vectors and the normal to the shock front. \citet{DeGasperin2014} suggested that this could be caused by the southern relic lying further away from us. In this scenario the radio emission we observe would have to pass through more magnetised, ionised plasma and therefore be subject to more Faraday rotation. However, spectroscopic observations of \target{} show that the redshift distributions of the member galaxies are well fit by a single Gaussian \citep{Golovich2019a}. This makes it unlikely that significant additional Faraday rotation due to projection effects is causing the observed difference in polarisation angle.

An alternate reason for the lack of polarised emission in the southern relic could be that turbulence in the relic has mixed the magnetic field lines. Turbulence mixes field lines at scales larger than the Alfv\'en scale ($l_A$), i.e. the scale at which the velocity of turbulence is equal to the Alfv\'en speed \citep[][]{Brunetti2016}. If this scale is smaller than the beam size, then we would observe no polarised emission. Simulations of cluster mergers show that merger shocks can generate turbulence, with $\leq$ 10\%  of the total kinetic energy flux dissipated into the generation of turbulence \citep[e.g.][]{Vazza2017}. The plasma that was (re-)accelerated by the merger shock which created the southern relic could instead have already been turbulent and the shock compression was insufficient to significantly alter its properties. The presence of turbulent motions within the relic may also help explain the irregular spectral values along the relic extension, with the presence of both steep ($\alpha \sim -1.2$) spectral indices close to the shock front and relatively flat ($\alpha \geq -1.0$) spectral index regions downstream (Fig.~\ref{fig:Spidxmaps}).

We can use the polarisation properties of the southern relic to constrain the turbulence within it, assuming the turbulence is generated by the merger shock. The fraction of kinetic energy flux converted to turbulent energy, $\eta$, is given by $\frac{1}{2} \eta \rho_u v_{s}^{3} \sim \rho_d \delta v_{0}^{2} v_d$, where $\rho_u$ and $\rho_d$ are the upstream and downstream ICM densities respectively, $v_s$ and $v_d$ the shock and downstream velocities and $\delta v_0$ the turbulent velocity at the injection scale. Turbulence needs time to decay from its injection scale, $L_0$, to $l_A$ before polarisation is removed. Since no polarisation is observed, even close to the shock edge, the decay time ($L_0/\delta v_0$) must be smaller than the time required to travel downstream by a distance of a few beams, ($b/v_d$), where $b$ represents a few beams. Assuming that the brightness substructures within the relic trace fluctuations in the shock structure caused by the turbulence and that the injection scale is approximately half of these fluctuations, we get that $L_0 \sim$ 100 kpc. An injection scale of 100 - 400 kpc is typical for the ICM \citep[][]{Brunetti2014}. Given that the beam size is $\sim$ 50kpc and the Mach number $\mach = 2.3$, we estimate that the efficiency of dissipating shock kinetic energy into turbulent energy exceeds 4\% at scales below 100 kpc. The stark differences between the north and south relics may be explained if the turbulence in the north relic is generated with lower efficiency, or at larger scales, than in the south relic. The differences may instead suggest that the Connection has played a role in the presence of turbulence in the downstream region of the southern relic (see Section~\ref{sec:S radio relic}).
This could also explain the irregular spectral index distribution in the southern relic (Fig.~\ref{fig:Spidxmaps}).

Some radio relics have been observed with such low polarisation fraction. \citet{Bonafede2009} reported a mean polarisation fraction of 8\% at 1.365 GHz in MACS J0717.5+3745, with regions of strong depolarisation. The polarisation structure of the western relic in Abell 3376 \citep[][]{Kale2012} shows a striking similarity to that of the southern relic in \target{}. \citet{Kale2012} reported patchy polarised emission, with electric field vectors not aligned normal to the shock front and suggested that this could be due to turbulence in the backflow of the shock.

The localised nature of the polarised emission and unexpected electric field vector orientation could instead indicate that we are observing emission from a polarised radio galaxy in projection within the relic. However, the area of polarised emission coincides with the brightest part of the radio relic, making it almost impossible to determine from the radio observations if there is indeed a radio galaxy producing the polarised emission we observe. We do not observe any likely optical counterparts in the Subaru image. From examination of the polarisation image, we find that there is low polarisation fraction (<10\%) across the entire relic, but below 3 $\sigma$. The localised emission seen in Fig.~\ref{fig:Polarisation Fraction} is therefore likely a consequence of lying at the brightest part of the relic.

The low signal to noise of the western side of the radio relic does not allow us to analyse the differences in polarisation structure between the two sides of the relic. Due to the complex nature of the southern relic (see  Section~\ref{sec:Connection}) it is difficult to draw any firm conclusions from the unusual polarisation structure.

\subsection{Southern Radio Relic Connection}
\label{sec:S radio relic}

Radio relics have previously been found with non-uniform brightness. For example, in the Toothbrush \citep{Rajpurohit2018} and Sausage \citep{Gennaro2018} galaxy clusters, the radio relics have bright filamentary structures. It is not certain what causes these structures, but they have been suggested to be the result of non-homogeneous mach numbers, magnetic field strengths or, in the re-acceleration scenario, fossil plasma distribution. At the resolution of our images, in the southern relic of \target{} we do not see filamentary sub-structures but a distinct east-west brightness jump. Instead, lower resolution images show diffuse radio emission expanding from the radio source S1 and bridging the gap to the southern relic (see Fig.~\ref{fig:Radio_images}). Extrapolating the lines drawn by the edges of the Connection to the radio relic front shows how the Connection aligns well with the brighter eastern side of the relic as well as with the edge of the brighter part of the relic on the west side (Fig.~\ref{fig:Radio optical labelled}).

One possibility is that we are observing two distinct shocks (labelled E and W in Fig.~\ref{fig:Radio optical labelled}) which only appear to be connected by projection effects. This is however unlikely as it would require very specific geometry for perfect alignment with the relic brightness jump to occur. Furthermore there are multiple pieces of evidence suggesting that this is indeed one single shock front, including the lack of a spectral index jump and similar spectral index distribution as a function of distance from shock front (Fig.~\ref{fig:Shock_Spidx_One_Graph}).
\citet{DeGasperin2014} reported that a changing spectral index along a radio relic may be indicative of a changing Mach number. If the southern relic in \target{} is in fact composed of two shocks superimposed on each other we might expect the shocks to have different Mach numbers and therefore a sharp change in spectral index at the border of the two shocks. Fig.~\ref{fig:Spidxmaps} shows a continuous spectral index across the brightness jump, with no sharp changes. Furthermore, the Mach numbers derived separately for the east and west sides of the radio relic agree, suggesting the southern relic can be explained by a single shock (Table~\ref{tab:Shock_props}). Although it is also possible that the southern relic is composed of two shocks, each with the same Mach number. 
Finally, the spectral index change towards the cluster centre is remarkably similar in the east and west sides of the relic (Fig.~\ref{fig:Shock_Spidx_One_Graph}). If the magnetic field strength was significantly greater on the east than on the west we would expect the synchrotron losses, and therefore the steepening of the spectral index, downstream of the shock to be greater in the east than the west. However, Inverse Compton scattering of cosmic microwave background (CMB) photons causes additional energy losses in the downstream region and therefore affects the spectral steepening. The equivalent magnetic field for these interactions is $B_\textrm{IC} = 3.2\mu G(1+z)^2$ \citep[][]{Longair2010}, giving $B_\textrm{IC} = 5\mu G$ at the redshift of \target{}. The magnetic field strengths measured in galaxy clusters are typically $\sim \mu G$ \citep[][and references therein]{VanWeeren2019}, so we would expect synchrotron losses to be significant.

The similarities of the two sides of the radio relic suggest therefore that something may have altered the properties of the relic on the eastern side but not on the west. The most obvious candidate is the bright discrete source close to the southern radio relic, labelled S1 in Fig.~\ref{fig:Radio optical labelled}. Using the catalogue of \citet{Golovich2019a} we find a cluster-member galaxy ($z = 0.304$) at the position of S1. It is plausible that we are observing sub-GeV electrons originating from the radio galaxy, which have been re-energised by the passing merger shock. There is precedent for such re-acceleration. For example, in Abell 3411-3412 the tail of a radio galaxy is connected to a radio relic \citep{VanWeeren2017}. The spectral index steepens along the tail before flattening out again towards the relic, suggesting that the shock wave has re-accelerated particles in the tail. In \target{} the integrated spectral index (Fig.~\ref{fig:Shock_Spidx_One_Graph}) steepens downstream of the shock front, before flattening again towards S1. This appears to suggest a similar scenario as in Abell 3411-3412. However, the spectral index maps (Fig.~\ref{fig:Spidxmaps}) are not so clear. The parts of the Connection visible in the spectral index maps do not show a clear steepening away from S1. The spectral steepening seen in the integrated spectral index plot is likely driven by the strong surface brightness decrease in the VLA map. This decrease creates a hole in the spectral index map where the VLA flux density is below $3\sigma_\textrm{rms}$ and is therefore excluded from local spectral index calculation (see Section~\ref{sec:Spectral Index}). However, this region is detected in our LOFAR images, so contributes significantly to the integrated flux density at 140 MHz. By measuring the flux density with LOFAR we calculate an upper limit of the spectral index within the hole to be $\alpha < -1.9$. The general spectral index steepening from S1 towards the northern edge of the relic suggests that the shock wave has passed through a radio galaxy tail, as in Abell 3411-3412. However, the complex spectral index profile makes interpretation challenging.
The coincidence of the Connection with a strong decrease in surface brightness (Fig.~\ref{fig:SRelic_SB_Profile}) and the profile of the integrated spectral index along the southern relic and Connection (Fig.~\ref{fig:Shock_Spidx_One_Graph}, red line) suggest a scenario in which the merger shock has re-accelerated sub-GeV electrons on the eastern side, whereas the west is produced by standard DSA, i.e. shock acceleration of electrons from the thermal pool. However, recent simulations by \citet{Zuhone2020} show that a merger-driven shock wave passing through the jets of an Active Galactic Nuclei (AGN) can produce an inhomogeneous cosmic ray fraction along the major axis of a shock. If this is the case in \target{}, this may explain the observed radio surface brightness discontinuity and therefore suggest that the east and west sides of the southern relic are generated by the same acceleration mechanism.

An alternate possibility is that the Connection is simply a product of projection effects and is in fact unrelated to S1 and instead turbulent radio halo emission, although this would not explain the surface brightness jump, polarisation and spectral index properties of the southern relic. Previous studies, for example in MACSJ1752.1+4440 \citep{Bonafede2012}, RXC J1314.4-2515 \citep[][]{Stuardi2019} and the Toothbrush \citep{VanWeeren2016a} suggest that in some cases there can be a connection between a merging cluster's giant radio halo and its radio relics. We investigate the possible detection of a giant radio halo in \target{} in the next section.

Within standard DSA the surface brightness jump could be explained if the brighter parts of the relic mark where the merger shock has interacted with higher density upstream regions. Following the analytical model of \citet{Hoeft2007}, the radio power generated by standard DSA is proportional to the electron density in the shock region. If we assume that the magnetic field strength and temperature across the southern relic are constant, the eastern side of the relic would require an electron density four times greater than the western side. Such a scenario would not explain the unusual polarisation structure observed in the southern relic. 

\subsection{Candidate Halo}
\label{sec:Halo}
\citet{DeGasperin2014} reported low-significance diffuse emission filling some of the ICM between the two relics in \target{}. They suggested that the diffuse emission could be from a radio halo. In our low resolution LOFAR images (Fig.~\ref{fig:Shock_Spidx_One_Graph}) we observe significant diffuse emission only on the eastern side of \target{}, coincident with the brightest region of the candidate halo in \citet{DeGasperin2014} (labelled "Tail" in Fig.~\ref{fig:Radio optical labelled}). The emission we see in LOFAR connects to $uv$-subtracted discrete sources (and a cluster-member galaxy, Fig.~\ref{fig:Radio optical labelled}). This may suggest that the emission is of galactic origin, or a blend of diffuse emission associated with the galaxies. Within just this tail of emission we measure a flux of $7 \pm 1$ mJy with LOFAR, but we do not detect significant emission with VLA. We can therefore set an upper limit on the spectral index of $\alpha < -1.3$, suggesting that this is emission from fossil plasma which has become invisible at higher frequencies due to synchrotron and IC losses.

Low resolution VLA images (Fig.~\ref{fig:Extraction_Regions}) show a low significance ($\leq 2\sigma_\textrm{rms}$) excess at the cluster centre, as reported in \citet{DeGasperin2014}. We therefore investigate the possibility that there is a radio halo in \target{}.

The low-mass end of radio halo correlations, for example radio halo power vs. mass, \citep[][]{Cassano2013,Cuciti2021} is relatively unexplored, owing to difficulty detecting radio halos in systems with comparatively low energy budgets available to inject large-scale cluster turbulence into the ICM. Detecting radio halos in low-mass systems and extending the existing studies on the correlation is therefore a crucial step forward and is indeed one of the goals of LOFAR \citep[e.g.][]{Cassano2012}.
If we measure the flux of the entire region between the two relics, avoiding S1, S2 and the Connection, we find $11 \pm 3$ and $0.6 \pm 0.3$ mJy for LOFAR and VLA respectively. We note that when excluding the bright eastern tail of emission seen in LOFAR we do not get a statistically significant detection with either LOFAR or VLA. Our flux measurements indicate a nearly 4$\sigma$ detection of diffuse emission in LOFAR but only 2$\sigma$ with VLA. Using the 3$\sigma$ error from VLA we set an upper limit on the spectral index of candidate halo emission in \target{} of $\alpha <$ -1.1. This aligns with the findings of \citet{Giovannini2009}, who found that radio halos typically have spectral indices of -1.4 $< \alpha <$ -1.1. Our flux measurement with LOFAR corresponds to a power at 150 MHz of $P_{150\textrm{MHz}} = (3.1 \pm 0.7) \times 10^{24}$ W/Hz.

To compare the candidate halo in \target{} with confirmed halos found in other clusters we use the correlation of \citet{vanWeeren2020}, which gives the relation between radio halo power at 150 MHz and cluster mass. It should however be noted that the correlation is not well defined at such low masses, as there are only three clusters included in the sample with masses <$5 \times 10^{14} \Msun$.
The correlation in \citet{vanWeeren2020} was calculated using the $M_{500}$ values from \textit{Planck}. We therefore use the \textit{Planck}-derived mass \citep[][$M_{500} = (4.7 \pm 0.3) \times 10^{14} \Msun$]{Planck2016} of \target{} to compare with the correlation. Using this we find that our candidate radio halo lies within the errors of the correlation.

Additionally, the radio power of the candidate halo in \target{} that we measure is in line with the findings of \citet{Bonafede2017}. In this work they investigated the lack of giant radio halos in galaxy clusters which host a pair of radio relics, including \target{}. By injecting mock halos into the WSRT data from \citet{DeGasperin2014} they derived an upper limit on the radio halo power of $P_{1.4\textrm{GHz}} = 3.8 \times 10^{23}$ W/Hz. The flux density upper limit from our VLA measurement corresponds to a power of $P_{1.4\textrm{GHz}} = 1.8 \times 10^{23}$ W/Hz, i.e. a factor of $\sim$ 2 smaller than the upper limit of \citet{Bonafede2017}.

From our data we cannot exclude the possibility that \target{} hosts a radio halo at its centre. However, most of the emission seen in LOFAR is localised to a small region connecting to discrete radio sources that have been subtracted. Additionally, one of the discrete sources coincides with a confirmed cluster-member galaxy. The brightest part of the candidate halo in VLA is on the eastern side of the ICM, suggesting that it is instead connected with the tail of emission seen in LOFAR. There may also be contribution from emission associated with discrete sources which were not fully subtracted from the $uv$-data.

\section{Conclusions}
\label{sec:conclusions}
In this paper we present LOFAR HBA (140 MHz) and VLA (1.5 GHz) radio observations and X-ray \textit{Chandra} data of \target{}. With these multi-frequency observations we carry out spectral and polarisation analysis of the double radio relics in \target{}. Our new, high-sensitivity radio observations have revealed the presence of a patch of diffuse emission (Connection) connecting a radio galaxy (S1) with the southern relic. The Connection coincides with a region in the radio relic in which there is a factor of $\sim$4 increase in surface brightness. Spectral analysis reveals a complex spectral index profile. The integrated spectral index steepens along the Connection, away from S1, which appears to suggest re-acceleration of an AGN lobe. The radio surface brightness discontinuity across the relic extension may suggest a scenario in which electrons are accelerated from the thermal pool on the western side of the relic, but re-accelerated from a pre-existing population of sub-GeV electrons in the east. Alternatively, some simulations suggest that shock passage across an AGN lobe can naturally reproduce the surface brightness gradient.
Our other findings are as follows:
\begin{itemize}
\item We confirm the detection of a pair of diametrically opposed radio relics in \target{}. The relics are approximately equidistant from the cluster centre, $\sim$ 1.1 and 1.0 Mpc for the northern and southern relics respectively. The radio relics have LLS of $\sim$ 0.9 and 1.5 Mpc and a flux ratio of 1:3.5.
\item The spectral index across the radio relics steepens away from the shock fronts, as expected from synchrotron and IC losses. The spectral index steepens from $\alpha = -0.5$ and $\alpha = -0.6$ to $\alpha = -1.5$ and $\alpha = -2.0$, for the northern and southern relics respectively. By measuring the spectral index along the shock fronts we derive Mach numbers of $\mach = 2.5 \pm 0.2$ and $\mach = 2.3 \pm 0.2$. These are significantly lower than those derived from the integrated spectral index of both radio relics.
\item We were unable to detect evidence of a shock in the \textit{Chandra} X-ray at either of the positions of the radio relics. However, the observation may not be deep enough to detect a shock and the location of the southern relic at the edge of two CCD chips makes observation of a shock even more challenging.
\item Measurements of the polarised emission at 1.5 GHz detected by VLA reveal areas of up to 60\% linearly polarised emission in the northern relic. The polarisation vectors, corresponding to the electric field, lie approximately perpendicular to the shock front, as expected for shock compression of magnetic fields. In the southern relic only small patches, at the brightest points in the relic, show very weakly polarised emission ($\leq$ 20\%). The electric field vectors lie at an angle relative to the shock-normal. We suggest that this could be caused by turbulence in the southern relic. The differences in the north and south relics suggest that the Connection is playing a role in this turbulence.
\item Low resolution LOFAR images show significant diffuse emission on the eastern side of \target{}, between the radio relics. The emission is localised to a tail of emission connecting to discrete radio sources, one of which coincides with a spectroscopically-confirmed cluster-member galaxy. From the non-detection in VLA we set an upper limit on the spectral index at $\alpha < -1.3$. We suggest that this is most likely emission from previous activity of a radio galaxy.
\item We consider the possibility that this emission is instead a candidate radio halo, as previously suggested in \citet{DeGasperin2014}. We find that the 150 MHz radio power lies on the $P_{150\textrm{MHz}} - M_{500}$ correlation of \citet{vanWeeren2020}, using the \textit{Planck}-derived mass.
\end{itemize}

\appendix
\section{VLA Calibrator Sources}
\label{sec:VLA_cals}
\begin{table*}
 \centering
 \caption{VLA Calibrator Sources}
 \label{tab:VLA_cals}
 %\resizebox{\columnwidth}{!}{%
 \begin{tabular}{lccccc}
  \hline
  Observation Date & \multicolumn{4}{c}{Calibrators} \\
  &  \\
  \hline
  & Primary & Polarisation Angle & Instrumental Polarisation & Phase \\
  2016 Feb 01 & 3C286 & 3C286 & J1407+2827 & J2022+6136\\
  2015 Feb 05 & 3C286 & 3C286 & J1407+2827 & J2022+6136\\
  2015 Feb 02-03 & 3C147 & 3C138 & 3C147 & J2022+6136\\
  
  \hline
 \end{tabular}
\end{table*}
Multiple calibrator sources are observed by VLA to properly calibrate each observation (see Section~\ref{sec:VLA obs}). The primary, polarisation and phase calibrators for each observation used in this paper are listed in Table~\ref{tab:VLA_cals}.

\section{Spectral Index Error Maps}
\label{sec:spidxerrmaps}
\begin{figure}
    \centering
    \includegraphics[width=\columnwidth]{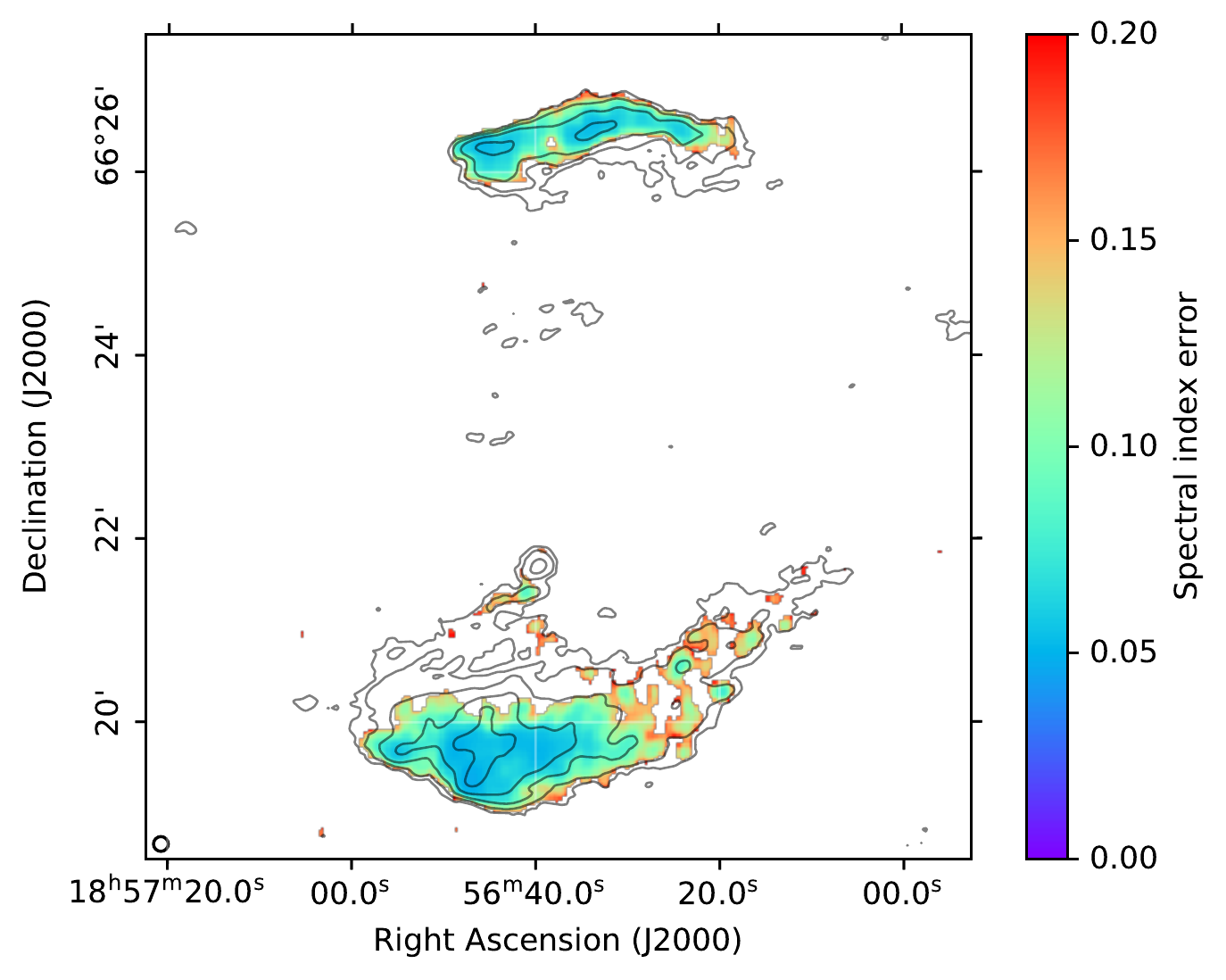}
    \includegraphics[width=\columnwidth]{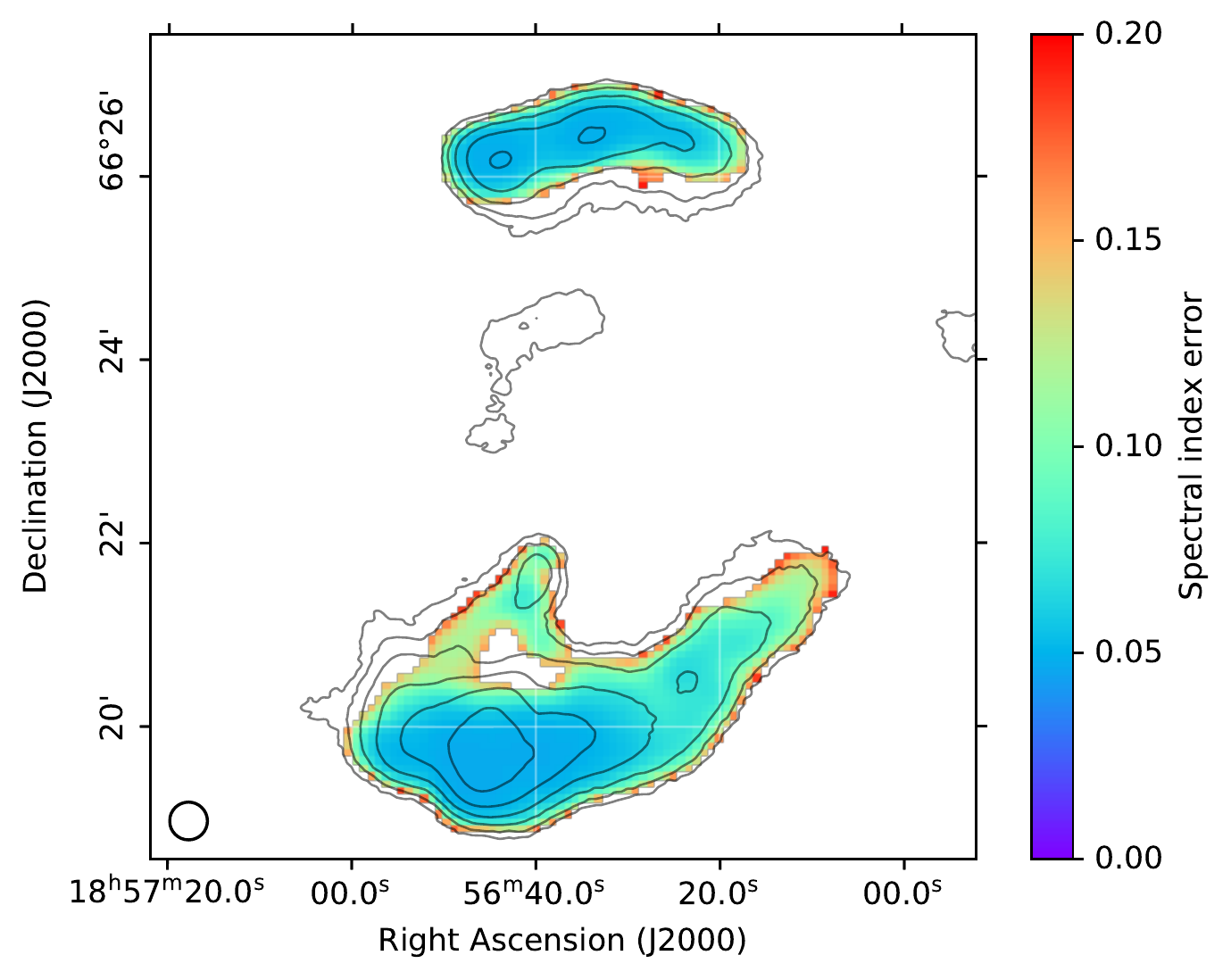}
    \caption{Spectral index error maps for Fig.~\ref{fig:Spidxmaps}.}
    \label{fig:Spidxerrmaps}
\end{figure}

The 140 - 1500 MHz spectral index was calculated for each pixel above 3$\sigma_{\rm{rms}}$ in both the LOFAR and VLA images. The resulting spectral index map is shown in Fig.~\ref{fig:Spidxmaps}. Fig.~\ref{fig:Spidxerrmaps} shows the corresponding spectral index error maps, calculated using Equation~\ref{eq:spidx err} for each pixel.

\section*{Acknowledgements}
FdG and MB acknowledge support from the Deutsche Forschungsgemeinschaft under Germany's Excellence Strategy - EXC 2121 ``Quantum Universe'' - 390833306.
VC acknowledges support from the Alexander von Humboldt Foundation. DNH acknowledges support from the ERC through the grant ERC-Stg DRANOEL n. 714245. AB acknowledges support from the VIDI research programme with project number 639.042.729, which is financed by the Netherlands Organisation for Scientific Research (NWO). WF, CJ and RK acknowledge support from the Smithsonian Institution and the Chandra High Resolution Camera Project through NASA contract NAS8-03060. RJvW acknowledges support from the ERC Starting Grant ClusterWeb 804208.
This paper is based (in part) on data obtained with the International LOFAR Telescope (ILT) under project code LC9\_036. LOFAR \citep[][]{VanHaarlem2013} is the Low Frequency Array designed and constructed by ASTRON. It has observing, data processing, and data storage facilities in several countries, that are owned by various parties (each with their own funding sources), and that are collectively operated by the ILT foundation under a joint scientific policy. The ILT resources have benefitted from the following recent major funding sources: CNRS-INSU, Observatoire de Paris and Universit\'e d'Orl\'eans, France; BMBF, MIWF-NRW, MPG, Germany; Science Foundation Ireland (SFI), Department of Business, Enterprise and Innovation (DBEI), Ireland; NWO, The Netherlands; The Science and Technology Facilities Council, UK. 
The National Radio Astronomy Observatory is a facility of the National Science Foundation operated under cooperative agreement by Associated Universities, Inc.
The scientific results reported in this article are based in part on observations made by the Chandra X-ray Observatory.

\section*{Data Availibility}
The data underlying this article will be shared on reasonable request to the corresponding author.

%%%%%%%%%%%%%%%%%%%% REFERENCES %%%%%%%%%%%%%%%%%%
\bibliographystyle{mnras}
\bibliography{references} 

% Don't change these lines
\bsp	% typesetting comment
\label{lastpage}
\end{document}